\documentclass[10pt,a4paper]{article}

\usepackage[english]{babel}
\usepackage{graphicx}
\usepackage{hyperref} 
\usepackage{geometry}
\usepackage{amsmath}
\usepackage{amssymb}
\usepackage{amsfonts}
\usepackage{multicol}
\usepackage{slashed}
\usepackage{float}
\usepackage{caption}
\usepackage{subcaption}
\usepackage{cite}
\usepackage{amsthm}
\usepackage{xcolor}
\usepackage{musicography}
\usepackage{bbm}
\usepackage{authblk}
\numberwithin{equation}{section}

\newcommand{\M}{\mathcal{M}}
\newcommand{\R}{\mathbb{R}}
\newcommand{\E}{\mathbb{E}}
\newcommand{\p}{\partial}
\newcommand{\rd}{\mathrm{d}}
\newcommand{\ri}{\mathrm{i}}

\title{\bf Quantum Theory, Gravity and Second Order Geometry}
\author[1]{Folkert~Kuipers\thanks{E-mail: f.kuipers@physik.uni-muenchen.de}}
\affil[1]{\em Arnold Sommerfeld Center for Theoretical Physics,\\
		Ludwig-Maximilians-Universit\"at M\"unchen, 80333 M\"unchen, Germany.}
\date{\today}

\begin{document}

\maketitle

\begin{abstract}
We argue that a consistent coupling of a quantum theory to gravity requires an extension of ordinary `first order' Riemannian geometry to second order Riemannian geometry, which incorporates both a line element and an area element.
This extension results in a misalignment between the dimension of the manifold and the dimension of the tangent spaces. In particular, we find that for a 4-dimensional spacetime, tangent spaces become 18-dimensional.
We then discuss the construction of physical theories within this framework, which involves the introduction of terms that are quadratic in derivatives in the action. On a flat spacetime, the quadratic sector is perpendicular to the first order sector and only affects the normalization of the path integral, whereas in a curved spacetime the quadratic sector couples to the first order sector. Moreover, we show that, despite the introduction of higher order derivatives, the Ostragradski instability can be avoided, due to an order mixing of the two sectors.
Finally, we comment on extensions to higher order geometry and on relations with non-commutative and generalized geometry.

\end{abstract}

\thispagestyle{empty}
\clearpage

\tableofcontents
\clearpage

\section{Introduction}\label{sec:Intro}

General Relativity lies at the heart of all our understanding of the gravitational force, and has been extremely successful as a description of classical gravity. However, issues start to arise when one subjects quantum theories to gravity. In this case, General Relativity no longer suffices for describing the gravitational force, which sparks the search for a theory that describes the interplay between quantum theory and gravity, and eventually for a theory of quantum  gravity.
\par 

From the mathematical perspective, the theory of General Relativity is formulated in the language of (pseudo-)Riemannian geometry, which thus serves as an indispensable tool for the description of classical gravity. Given this fact and the fact that the interplay between quantum theory and gravity is still only partially understood, one faces the question: does Riemannian geometry suffice for describing the interplay between quantum theory and gravity or is a generalization of classical Riemannian geometry to some notion of `quantum Riemannian geometry' necessary?
\par

The idea that Riemannian geometry may be generalized is not new, and can, in fact, be traced back to Riemann himself who wrote \cite{Riemann:1868,Jost:2016}:

\begin{quotation}
	``For space, when the position of points is expressed by the rectilinear coordinates, $ds=\sqrt{\sum(dx)^2}$; space is therefore included in this simplest case. The next case in simplicity includes those manifoldnesses in which the line-element may be expressed as the fourth root of a quartic differential expression. The investigation of this more general kind would require no really different principles, but would take considerable time and throw little new light on the theory, especially as the results cannot be geometrically expressed;''
\end{quotation}
Thus, Riemann already suggested to generalize the line element from
\begin{equation}\label{eq:LineElement}
	\rd s^2 = g_{\mu\nu} \, \rd x^\mu \rd x^\nu 
\end{equation}
to 
\begin{equation}
	\rd s^4 = G_{\mu\nu\rho\sigma} \, \rd x^\mu \rd x^\nu \rd x^\rho \rd x^\sigma \, .
\end{equation}
We note that, by grouping the $\rd x$ terms, this generalization can also be understood as an area element of the form
\begin{equation}\label{eq:AreaElement}
	\rd A^2 = G_{|\mu\nu|\rho\sigma|} \, \rd\langle x^\mu, x^\nu\rangle \, \rd\langle x^\rho, x^\sigma \rangle \, .
\end{equation}
As illustrated by this quote, Riemann did not delve into such extensions as it would take `considerable time' and `throw little new light on the theory'. As we will discuss in this work, from the perspective of classical General Relativity, such extensions are indeed of no interest. In fact, such extensions are incompatible with the basic assumption that spacetime looks locally like a Minkowski space. It must be noted, however, that Riemannian geometry was formulated several decades before the introduction of quantum mechanics and modern probability theory \cite{Kolmogorov}. Here, we will argue that only in light of these modern theories, such extensions become physically meaningful and can shine new light on these theories.
\par 

The suggestion that a consistent theory that incorporates both quantum theory and gravity requires a generalization beyond Riemannian geometry is not novel either. 
There exist various examples of `quantum geometry' frameworks, e.g. non-commutative geometry, that propose generalizations of Riemannian geometry. 
Such generalization are of particular interest for quantum gravity research, since they usually come with a small length scale, that quantifies the deformations from Riemannian geometry. This scale can then be taken as a cut-off, which could help to tame the divergences encountered in the quantization of gravity.
\par 

In this work, we discuss an example of a generalization of Riemannian geometry, which is another proposal for such a `quantum geometry', that incorporates both the line element \eqref{eq:LineElement} and the area element \eqref{eq:AreaElement}. More precisely, we will discuss second order geometry, which is a minimal extension of the usual `first order' Riemannian geometry to higher order geometry. This framework was originally proposed in the mathematics literature of the early 1980s by Schwartz \cite{Schwartz:1984} and Meyer \cite{Meyer:1981}. Their purpose for developing this generalization was in describing stochastic theories on manifolds. The advantage of this framework is that it allows to make the notion of general covariance manifest in stochastic theories.
\par

Exploiting the tight connection between quantum theory and stochastics that was established by the path integral formulation \cite{Feynman:1948ur,FKac,Glimm:1987ylb}, we will argue that second order geometry may equally well be used for the construction of manifestely covariant quantum theories on curved spacetime, which serves as a first step towards the construction of a theory of quantum gravity.\footnote{The issue of covariance in quantum field theory on curved spacetime has a long history that can be traced back to the 1950s, cf. e.g.   Refs.~\cite{DeWitt:1967ub,ChapelHill}. In this work we aim to provide a complementary perspective on this issue using the language of second order geometry.}
\par 

This paper is organized as follows: in section \ref{sec:PIProp}, we review how smooth worldlines are replaced by a set of rough paths in the path integral formulation; in section \ref{sec:Differentials}, we discuss how the lack of regularity modifies the differentials defined along these paths; in section \ref{sec:DiffOperators}, we show how this in turn modifies the differential forms; in section \ref{eq:2ndGeom}, we review the basics of second order geometry, which is a differential geometry based on these modified differential forms; in section \ref{sec:Metric}, we discuss the construction of a metric in second order geometry, which forms the basis of (pseudo-)Riemannian second order geometry; in section \ref{eq:PhysTheory}, we discuss how second order geometry leads to modifications of the physical theories that are built on top of the geometry; in section \ref{sec:Generalizations}, we discuss some generalizations beyond second order geometry; finally, in section \ref{sec:Conclusion}, we conclude. Furthermore, in appendix \ref{Sec:Martingales}, we provide some more background on the modified differential from section \ref{sec:Differentials}, and, in appendix \ref{ap:higherOrder}, we generalize some of the expressions from sections \ref{sec:Differentials} and \ref{sec:DiffOperators}. Throughout the work we use natural units $\hbar = c = k_B = G =1$, unless specified otherwise.

\section{From classical worldlines to quantum worldsheets}\label{sec:PIProp}
We consider a Riemannian manifold $\M$ of dimension $n\in\mathbb{N}$. Given a coordinate chart, a classical trajectory for a non-relativistic particle of mass $m>0$ is a function
\begin{equation}
	X : \mathcal{T} \rightarrow \M \qquad {\rm s.t. }\qquad t \mapsto \{X^i(t) \, |\, i\in\{1,...,n\}\}\, ,
\end{equation}
where $t\in\mathcal{T}\subseteq\R$ denotes the time parameter and $\mathcal{T}$ is a time interval. Similarly, for a relativistic particle with mass $m\geq0$, one considers a $n$-dimensional Lorentzian manifold $\M$. Then, trajectories are functions
\begin{equation}\label{eq:Classpath}
	X : \mathcal{T} \rightarrow \M \qquad {\rm s.t. }\qquad \tau \mapsto \{X^\mu(\tau) \, | \, \mu\in\{0,1,...,n-1\}\}\, ,
\end{equation}
where $\tau\in\mathcal{T}$ denotes the proper time parameter.
\par

Our goal is to generalize these classical trajectories to a notion of `quantum' trajectories. This generalization will apply to both non-relativistic and relativistic theories. However, we will predominantly use the notation of relativistic theories. Thus, in the remainder, we will denote the time parameter by $\tau$ and label the coordinates with Greek indices. The results for the non-relativistic theory can easily be obtained by replacing $x^\mu\rightarrow x^i$ and $\tau\rightarrow t$.
\par 

A crucial assumption in the study of classical trajectories is the fact that they are continuously differentiable.\footnote{In fact, classical mechanics requires the existence of a well-defined acceleration. Therefore, classical trajectories are at least twice continuously differentiable. For simplicity, they are often assumed to be smooth.} Thus, the following limits
\begin{align}\label{eq:DefClassVel}
	v_+(\tau) &= \lim_{d\tau\rightarrow 0} \frac{x(\tau+d\tau) - x(\tau)}{d\tau} \, , \nonumber\\
	v_-(\tau) &= \lim_{d\tau\rightarrow 0} \frac{x(\tau) - x(\tau-d\tau)}{d\tau}
\end{align}
exist\footnote{If $\mathcal{T}$ is an open interval $(a,b)$ with $a,b\in\mathbb{R}$, $V_\pm$ exist at the boundaries, but may tend to $\pm\infty$. Nevertheless, on the interval itself the limits exist and are finite. A similar reasoning applies to the half open intervals $(a,b]$ and $[a,b)$.} and are such that\footnote{There are counterexamples to this statement: for example, this equality fails for a classical particle in a box, where the two limits do not coincide, whenever the particle hits the boundaries of the box. However, in these examples, the limits \eqref{eq:DefClassVel} still exist and the number of discontinuities where \eqref{eq:VelLRlimEq} is violated are countable, such that the times $\tau_k\in\mathcal{T}$ at which the velocity is discontinuous are isolated.}
\begin{equation}\label{eq:VelLRlimEq}
	v(\tau) := v_+(\tau) = v_-(\tau) \, .
\end{equation}
Moreover, for any higher power one obtains
\begin{equation}
	\lim_{d\tau \rightarrow 0 } = \frac{[x(\tau+d\tau) - x(\tau)]^k}{d\tau} = \lim_{d\tau\rightarrow 0} O(d\tau^{k-1}) = 0 \qquad \forall\, k>1 \, .
\end{equation}
\par

In quantum mechanics the position $X(\tau)$ is replaced by an operator $\hat{x}(\tau)$, which is time dependent in the Heisenberg picture. For these operators the limits \eqref{eq:DefClassVel} are ill-defined. However, if one first takes an expectation value before taking the limit, the limits become well-defined, but they may be unequal, i.e. $v_+\neq v_-$.
Furthermore, when one calculates higher order moments one finds that
\begin{equation}\label{eq:Quant2Vel}
	\lim_{d\tau\rightarrow 0} \left\langle \frac{ [\hat{x}(\tau+d\tau) - \hat{x}(\tau)]^2}{d\tau} \right\rangle \neq 0\, ,
\end{equation}
whereas
\begin{equation}
	\lim_{d\tau\rightarrow 0} \left\langle \frac{ [\hat{x}(\tau+d\tau) - \hat{x}(\tau)]^k}{d\tau} \right\rangle =0  \qquad \forall k>2 \, .
\end{equation}
In fact, the non-vanishing of the limit~\eqref{eq:Quant2Vel} is essential for deriving the commutation relation from the path integral formulation, and thus a necessary condition in establishing equivalence between the path integral formulation and canonical quantization. More precisely, in non-relativistic quantum mechanics\footnote{This expression can be generalized to non-relativistic quantum mechanics on curved spacetime, where the right hand side becomes $\frac{\ri\, \hbar}{m}g^{\mu\nu}$ and the commutation relation is $[\hat{x}^\mu,\hat{p}_\nu]=\ri \, \hbar \, \delta^\mu_\nu$, cf. e.g. \cite{Reisenberger:2001pk,Giovannetti:2022pab,Kuipers:2023pzm,Kuipers:2023ibv}.} on $\R^n$ one requires\footnote{Cf. section 9 in Ref.~\cite{Feynman:1948ur}.}
\begin{equation}\label{eq:conditionFeynman}
	\lim_{d\tau\rightarrow 0} \left\langle \frac{[\hat{x}^i(\tau+d\tau) - \hat{x}^i(\tau)] [\hat{x}^j(\tau+d\tau) - \hat{x}^j(\tau)]}{d\tau} \right\rangle
	= \frac{\ri \, \hbar}{m} \, \delta^{ij} \, ,
\end{equation}
where $m$ is the mass of the particle, which induces the commutation relation
\begin{equation}\label{eq:CommutatorCanonical}
	m \, \delta_{jk}\, [\hat{x}^i,\hat{v}^k] = [\hat{x}^i,\hat{p}_j] = \ri \, \hbar \, \delta^i_j \, . 
\end{equation}
\par 

We will now make this observation more precise. According to the path integral formulation, the classical trajectories must be replaced by a set of `quantum trajectories' that can be drawn from some sample space $\Omega$. Thus, the trajectories are now functions
\begin{equation}\label{eq:ClasspathRLT}
	X : \Omega \times \mathcal{T} \rightarrow \M \qquad {\rm s.t. }\qquad (\omega,\tau) \mapsto X(\omega,\tau)\, .
\end{equation}
Then, for any $\omega\in\Omega$, one obtains a unique trajectory, known as a sample path, given by
\begin{equation}
	X(\omega,\cdot) : \mathcal{T} \rightarrow \M \qquad {\rm s.t. }\qquad \tau \mapsto X(\omega,\tau)\, .
\end{equation}
These paths are selected according to a probability measure
\begin{equation}
	\mathbb{P}:\Sigma(\Omega)\rightarrow[0,1] \, ,
\end{equation} 
where $\Sigma$ is a sigma algebra over $\Omega$. The observables of the quantum theory can be calculated by averaging over all paths weighted by this probability measure. This can be done by constructing the probability amplitude of transforming from a state $x_0$  at $\tau_0$ to a state $x_f$ at $\tau_f$. This probability amplitude is given by
\begin{equation}\label{eq:Pathint}
	%\langle x_f \, | \,  e^{- \frac{\ri}{\hbar} H \, (\tau_f -\tau_0)}\, |\, x_0 \rangle  =
	\langle x_f, \tau_f \, | \, x_0, \tau_0 \rangle
	%= W(x_f,\tau_f;x_0,\tau_0) 
	= \int_{X(\tau_0)=x_0}^{X(\tau_f)=x_f} e^{\ri \, S(X)} \, DX \,.
\end{equation}
The right hand side is called the path integral and its domain is the space of all paths $X(\tau)$, between the fixed endpoints $(\tau_0,x_0)$ and $(\tau_f,x_f)$ that are weighed by the `probability density' 
\begin{equation}\label{eq:PathIntMeasure}
	\rho(X) \sim e^{\ri \, S(X)} = e^{\ri \int_\mathcal{T} L(X,\dot{X},\tau) d\tau} \, .
\end{equation}
All $n$-point functions related to the position of the particle can then be calculated by constructing the characteristic functional
\begin{align}\label{eq:CharacteristicFunctional}
	\phi_X(J) 
	&= \E\left[ e^{\ri \, \langle J , X\rangle} \right] \nonumber\\
	&= \int_{\Omega} e^{\ri \, \langle J , X\rangle} \, \rd\mathbb{P}(\omega) 
	\nonumber\\
	&= \int_{L^2(\Omega)} e^{\ri \, \langle J , X\rangle} \, \rd\mu(X) \nonumber\\
	&= \frac{\int_{X(\tau_0)=x_0}^{X(\tau_f)=x_f} e^{\ri S(X) + \ri \langle J , X\rangle}  \, D X}{\int_{X(\tau_0)=x_0}^{X(\tau_f)=x_f} e^{\ri S(X)}  \, D X} 
	\nonumber\\
	&= \frac{\left\langle x_f , \tau_f \, \big| \, e^{\ri \langle J , X\rangle} \, \big| \, x_0,\tau_0 \right\rangle }{\langle x_f,\tau_f\, |\,  x_0,\tau_0 \rangle }
	\, ,
\end{align}
%
%\int_\mathcal{T} \langle J , X\rangle(\tau) \, d\tau
where $\mu= \mathbb{P}\circ X^{-1}$ is the probability measure on $L^2(\Omega)$ and $\langle J, X \rangle = \int_\mathcal{T} J_\mu(\tau) \, X^\mu(\tau) \, d\tau$. Note that, in the first three lines, we used the standard definitions of the characteristic functional and expectation value from probability theory, where expectation values are commonly denoted by $\E[]$. In the fourth line, we equate this expression to the path integral formulation of quantum mechanics, where expectation values are commonly denoted by the angle brackets $\langle\rangle$.\footnote{In the remainder of the text, we will stick to these conventions:  the conditional expectation of $X$ given $Y$ will be denoted by $\E[X|Y]$ when defined within Kolmogorov's axioms of probability theory, whereas $\langle X \rangle_Y$ is used to denote expectation values when defined within Von-Neumann's axiomization of quantum theory and to denote ensemble averages in a statistical theory.}
\par 

A major issue in the path integral formulation of quantum mechanics is its existence. Nevertheless, assuming its existence, it can serve as a powerful tool in calculating observables of quantum theories. A major issue, in proving the existence of path integrals, is that the probability measure $\mathbb{P}$ does not exist \cite{Cameron,Daletskii}.\footnote{If $\mathbb{P}$ does not exist, the first three lines in eq.~\eqref{eq:CharacteristicFunctional} are ill-defined. Thus, one can not use this expression to provide a mathematically rigorous definition of the path integral. Instead, one has to resort to others ways of making mathematical sense of the path integral, cf. e.g. Ref.~\cite{Albeverio}.}
However, this problem can be ameliorated using the Wick rotation \cite{Wick:1954eu}, which changes the theory in two ways: 
\begin{itemize}
	\item The probability amplitude $e^{\ri S(X)}$ is turned into $e^{-S(X)}$.
	\item If a relativistic theory is studied, the signature of the manifold is changed from Lorentzian to Euclidean, such that for a Lorentzian theory $S$ is turned into $S_{\rm Eucl}$.
\end{itemize}
Due to the second property, the original non-Wick rotated theory is often referred to as the Lorentzian theory, whereas the Wick rotated theory is called the Euclidean theory. In both the relativistic and non-relativistic theory, the path integral \eqref{eq:Pathint} becomes a well-defined object after the Wick rotation. The associated measure
\begin{equation}
	\rd\mu(X) = e^{- S_{\rm Eucl}(X)} \, D X
\end{equation}
is a Wiener measure and $X$ is a Wiener process with drift \cite{Wiener,FKac}.\footnote{This process is also known as Brownian motion, as it can be used to describe Brownian motion.}
\par 

The fact that the Euclidean path integral is related to the Wiener process can be put to good use. In particular, this allows to derive properties of the paths that one integrates over in the path integral formulation. For example, we have the following properties, cf. e.g. \cite{Morters:2010}:
\begin{itemize}
	\item The paths are almost surely\footnote{Meaning that the set of paths $A\subseteq\Omega$ for which this statement holds has probability $\mathbb{P}(A)=1$.} continuous, but nowhere differentiable. More precisely, they are almost surely $\alpha$-H\"older continuous for $\alpha<\frac{1}{2}$, but\footnote{Points where the paths are $\alpha$-H\"older continuous for $\alpha=\frac{1}{2}$ exist almost surely, but they are rare.} not for $\alpha>\frac{1}{2}$.
	Thus, the path integral replaces classical paths $X\in\mathcal{C}^2(\mathcal{T},\M)$ with `quantum paths' $X(\omega)\in C^{1/2}(\mathcal{T},\M)$.
	\item The Hausdorff dimension of the range of $X$ is given by
	\begin{equation}
		{\rm dim_H} \Big(\Big\{X(\omega,\tau) \, \Big|\, \tau\in\mathcal{T} , \, \omega\in\Omega\Big\}\Big) 
		= 	
		\begin{cases}
			1 \qquad {\rm if} \quad n=1 \, ,\\
			2 \qquad {\rm if} \quad n\geq 2 \, .
		\end{cases}
	\end{equation}
	Thus, the path integral replaces classical worldlines with worldsheets.
\end{itemize}
We note that these properties are proved for the Euclidean theory. However, there is no indication that these properties are altered by the Wick rotation. In fact, the path integral of the single particle as described above can be described using two maximally correlated Wiener processes \cite{Kuipers:2023pzm,Kuipers:2023ibv}, which suggests that the properties are indeed preserved by the Wick rotation.

\section{Differentials along non-differentiable paths}\label{sec:Differentials}
As discussed in the previous section \ref{sec:PIProp}, the paths that a path integral sums over are of the type\footnote{Strictly speaking, one only has $X\in \mathcal{C}^{\alpha}(\mathcal{T},\M)$ only for $\alpha\in[0,1/2)$.} $X\in \mathcal{C}^{1/2}(\mathcal{T},\M)$. Let us define the forward and backward differentials along such paths:
\begin{align}\label{eq:DefDiffpm}
	\rd_+X(\tau) := X(\tau + d\tau) - X(\tau)\, ,\nonumber\\
	\rd_-X(\tau) := X(\tau) - X(\tau-d\tau) \, ,
\end{align}
where we assume $0<d\tau\ll1$. Then, the H\"older regularity of the paths implies that
\begin{equation}
	\rd_\pm X(\tau) = O(d\tau^{1/2}) \, ,
\end{equation}
such that
\begin{equation}\label{eq:DivVel}
	\lim_{d\tau \rightarrow 0} \left|\frac{\rd_\pm X(\tau)}{d\tau} \right| = \infty \, .
\end{equation}
Thus, we cannot define a velocity along the paths, as we did for the classical paths in eq.~\eqref{eq:DefClassVel}. Nevertheless, using the relation between the path integral and the Wiener integral, and the fact that for the Wiener process the limit becomes well-defined, when an expectation value is introduced, cf. e.g. \cite{Nelson:1967}, we can write 
\begin{align}\label{eq:VelField}
	v_\pm^\mu(x,\tau) &:= \lim_{d\tau \rightarrow 0} \E\left[\frac{\rd_\pm X^\mu(\tau)}{d\tau} \, \Big| \, X(\tau)=x \right] ,
\end{align}
where $v_+\neq v_-$. Additionally, there exist well-defined fields given by\footnote{The object $\rd X \rd X$ is called quadratic variation. In the stochastics literature, it is usually denoted by square and angle brackets: $ v_{2\pm} \, d\tau = \rd_\pm\langle X, X\rangle = \E\left[\rd_{\pm}[ X , X]\, | \, X \right]$.}
\begin{align}\label{eq:Vel2Initial}
	v_{2\pm}^{\mu\nu}(x,\tau) 
	&:= \lim_{d\tau \rightarrow 0} \E\left[\pm \frac{\rd_{\pm} X^\mu(\tau) \, \rd_{\pm} X^\nu(\tau) }{d\tau} \, \Big| \, X(\tau)=x \right],
\end{align}
where $v_\pm$ has the dimension of velocity, i.e. $[v_\pm] = L/T$, whereas $[v_{2\pm}] = L^2/T$.
This suggests that the divergence, found in eq.~\eqref{eq:DivVel}, may be isolated from the finite part, calculated in \eqref{eq:VelField}, such that we can write\footnote{If one defines $b_\pm = \pm \frac{\sqrt{\pm 1}}{(1/2)!} \tilde{b}_\pm = \frac{2\, \tilde{b}_\pm}{\sqrt{\pm \pi}} $, the expansion mimicks a Taylor expansion that includes fractional derivatives $\tilde{b}= \frac{d^{1/2}X}{d\tau^{1/2}}$. For simplicity of notation, we will not include such factors here.}
\begin{equation}\label{eq:DifferentialImposed}
	\rd_\pm X^\mu(\tau) = b_\pm^\mu(X(\tau),\tau) \, d\tau^{1/2} + v_\pm^\mu(X(\tau),\tau) \, d\tau + o(d\tau) \, .
\end{equation}
We emphasize that this is only a formal expression, as we have not specified the mathematical properties of the objects $b_\pm$ and $v_\pm$. This will be remedied in section \ref{eq:2ndGeom}, where a more in-depth discussion of the objects $v,v_2$ is given, and in appendix \ref{Sec:Martingales}, were the objects $b_\pm$ are discussed in more detail. 
\par 

Using the expression \eqref{eq:DifferentialImposed} and the `velocities' \eqref{eq:VelField} and \eqref{eq:Vel2Initial}, one finds that the first and second moment of $b_\pm$ are given by\footnote{Note that in the context of quantum theory, one expects the term at order $d\tau^{1/2}$ to vanish in the limit $\hbar\rightarrow 0$. This suggests to define $b_\pm = \sqrt{\hbar/\mu} \, \tilde{b}_\pm$, where $\mu$ is a mass scale and $\tilde{b}_\pm$ is dimensionless. Considering eq.~\eqref{eq:conditionFeynman}, one would then fix $\mu=m$, which is the only mass scale that is currently present. For simplicity of notation, we will not make such dimensional quantities explicit here.}
\begin{align}
	\E\left[ b_\pm^\mu \, \Big| \, X(\tau)=x \right] &= 0 \, ,\\
	\E\left[ \pm\, b_\pm^\mu \, b_\pm^\nu \, \Big| \, X(\tau)=x \right] &=  v_{2\pm}^{\mu\nu}(x,\tau) \, .\label{eq:Vel2B}
\end{align}
Moreover, higher order moments are calculated using Isserliss' theorem \cite{Isserlis:1918}:
\begin{equation}
	\E\left[\prod_{i=1}^k b_\pm^{\mu_i}\right] 
	= \sum_{i=1}^{k-1} \pm v_{2\pm}^{\mu_i\mu_k} \, \E\left[\prod_{j=1: j\neq i}^{k-1} b_\pm^{\mu_j}\right]\, .
\end{equation}
Products involving both $b_+$ and $b_-$ may also be considered, which leads to the construction of two more `velocities':
\begin{align}
	v_{+-}^{\mu\nu} &= \E\left[ b_+^\mu \, b_-^\nu \, \Big| \, X(\tau)=x \right] , 
	\nonumber\\
	v_{-+}^{\mu\nu} &= \E\left[ b_-^\mu \, b_+^\nu \, \Big| \, X(\tau)=x \right] .
\end{align}
As before, these products can be generalized to higher order powers involving both $b_+$ and $b_-$. However, since $b_+,b_-$ may not commute, these should be calculated using Wick's theorem \cite{Wick:1950ee}.
\par 

We may generalize the definition of the forward and backward differentials \eqref{eq:DefDiffpm} to a family
\begin{equation}\label{eq:GenDifferential}
	\rd_a X(\tau) := X\Big(\tau + \frac{1+a}{2}\, d\tau\Big) - X\Big(\tau - \frac{1-a}{2}\, d\tau\Big)\, ,
\end{equation}
where $a\in[-1,1]$. Using eqs.~\eqref{eq:DefDiffpm} and \eqref{eq:DifferentialImposed}, this expression can be evaluated as
\begin{align}
	\rd_a X(\tau) 
	&= X\Big(\tau + \frac{1+a}{2}\, d\tau\Big) - X(\tau) + X(\tau) - X\Big(\tau - \frac{1-a}{2}\, d\tau\Big) \nonumber\\
	&= \frac{\sqrt{1+ a} \, b_+ + \sqrt{1-a} \, b_-}{\sqrt{2}} \, d\tau^{1/2} + \frac{(1+a)\, v_+ + (1-a)\, v_- }{2} \,  d\tau + o(d\tau) \,.
\end{align}
As we will be particularly interested in the cases $a\in\{-1,0,1\}$, we will evaluate this expression for the case $a=0$, which yields
\begin{equation}
	\rd_0 X(\tau)
	= \frac{b_+ + b_-}{\sqrt{2}} \, d\tau^{1/2} + v_\circ \, d\tau + o(d\tau) \,,
\end{equation}
where we defined
\begin{align}
	v_\circ &:= \frac{v_+ + v_-}{2} \, .
\end{align}
\par 

We can also consider second order differentials that are given by
\begin{align}\label{eq:AccelerationGen}
	\rd^2_{a} X(\tau)
	&:= \rd_{a} \rd_{a} X(\tau) 
	\nonumber\\
	&= X[\tau + (a+1)\, d\tau ] 
	- 2 \, X[ \tau + a\, d\tau]
	+ X[\tau + (a-1)\,d\tau] \, .
\end{align}
This expression can be evaluated, yielding
\begin{equation}
	\rd_a^2 X 
	=
	\begin{cases}
		\left[\sqrt{1+a} \, b_+ - (\sqrt{1-a} - 2\, \sqrt{-a} ) \, b_- \right] d\tau^{1/2} + (1+a)\, (v_+-v_-) \, d\tau + o(d\tau)\, , \qquad &{\rm if} \quad a<0 \, ;\\
		( b_+ - b_- ) \, d\tau^{1/2} + (v_+-v_-) \, d\tau + o(d\tau) \, , \qquad &{\rm if} \quad a=0 \, ;\\
		\left[(\sqrt{1+a} - 2 \, \sqrt{a} )\, b_+ - \sqrt{1-a} \, b_- \right] d\tau^{1/2} + (1-a) \, (v_+-v_-) \, d\tau + o(d\tau)\, , \qquad &{\rm if} \quad a>0 \, .
	\end{cases}
\end{equation}
Thus, we see that the acceleration contains a part that appears at the same order in $d\tau$ as the velocities $v_\pm$, This part is given by
\begin{align}
	\lim_{d\tau \rightarrow 0} \E\left[\frac{\rd_a^2 X(\tau)}{ d\tau} \, \Big| \, X(\tau)=x \right] 
	&= 2 \, (1 - |a|) \, v_\perp \, ,
\end{align}
where
\begin{equation}
	v_\perp := \frac{v_+ - v_-}{2} \, .
\end{equation}
\par 

In addition, we can define another acceleration type operator of the form\footnote{By construction we have the commutation relation $[\rd_{a_1},\rd_{a_2}]=0\quad \forall\,a_1,a_2\in[-1,1]$.}
\begin{align}
	|\rd_a^2| X(\tau) 
	&:= \rd_a \rd_{-a} X(\tau)\nonumber\\
	&= X(\tau + d\tau) 
	- 2 \, X(\tau )
	+ X(\tau - d\tau)
	\nonumber\\
	&=
	( b_+ - b_- ) \, d\tau^{1/2} + (v_+-v_-) \, d\tau + o(d\tau)\, .
\end{align}
For this second order operator, one obtains
\begin{align}
	\lim_{d\tau \rightarrow 0} \E\left[\frac{|\rd_a^2| X(\tau)}{ d\tau} \, \Big| \, X(\tau)=x \right] 
	&= 2 \, v_\perp \, .
\end{align}
\par
 
One may generalize this analysis to higher order differentials $\rd_a^kX$ of any power $k\in\mathbb{N}$. The results of this generalization can be found in appendix \ref{Ap:HighOrderDiff}. Here, we only note that all terms at order $d\tau^{1/2}$ are given by linear combinations of $b_+,b_-$, and all terms at order $d\tau$ by linear combinations of $v_+,v_-$. Therefore, for trajectories $X\in C^{1/2}(\mathcal{T},\M)$, the spaces spanned by $b_+,b_-$ and $v_+,v_-$ contain all the information associated to the differentials. In contrast, for trajectories $X\in C^{1}(\mathcal{T},\M)$, all this information is stored in a single velocity vector $v$.
\par 

In the remainder of this work, we will focus on the cases $a\in\{-1,0,1\}$, which are the standard choices considered in the literature. In the mathematics literature, the case $a=0$ is related to the Stratonovich formulation of stochastic calculus, whereas the cases $a\in\{-1,1\}$ are related to the It\^o formulation. In the physics literature, on the other hand, the case $a=0$ is associated to the midpoint evaluation of the path integral, whereas the cases $a\in\{-1,1\}$ are associated to the prepoint and postpoint evaluation of path integrals, cf. e.g. Ref.~\cite{Chaichian:2001cz}.

\section{Differential operators}\label{sec:DiffOperators}
After having introduced the differentials $\rd_\pm X$ in the previous section \ref{sec:Differentials}, we may turn our attention to differential operators that act on functions $f\in C^2(\M,\R)$. Analogously to eq.~\eqref{eq:DefDiffpm}, we define, along any trajectory $X:\mathcal{T}\rightarrow\M$,
\begin{align}\label{eq:DiffOpFnct}
	\rd_+ f(X(\tau)) &:= f(X(\tau+d\tau)) - f(X(\tau)) \, ,\nonumber\\
	\rd_- f(X(\tau)) &:= f(X(\tau)) - f(X(\tau-d\tau)) \, .
\end{align}
Then, using the differentiability of $f$ and the results from previous section, we obtain
\begin{align}\label{eq:DiffFunc+}
	\rd_+ f(X) 
	&= f(X+\rd_+X) - f(X) \nonumber\\
	&= \p_\mu f(X) \, \rd_+X^\mu + \frac{1}{2} \p_\nu\p_\mu f(X) \, \rd_+ X^\mu \rd_+ X^\nu + o(d\tau) 
\end{align}
and
\begin{align}\label{eq:DiffFunc-}
	\rd_- f(X) 
	&= f(X) - f(X-\rd_-X) \nonumber\\
	&= \p_\mu f(X) \, \rd_-X^\mu - \frac{1}{2} \p_\nu\p_\mu f(X) \, \rd_- X^\mu \rd_- X^\nu + o(d\tau) \, .
\end{align}
As in eq.~\eqref{eq:GenDifferential}, we can also consider more general differential operators labeled by $a\in[-1,1]$. We are mainly interested in $a=0$ for which we obtain, cf. eq.~\eqref{Calc:d0f},
\begin{align}
	\rd_0 f(X)
	&:= f(X(\tau+d\tau/2)) -  f(X(\tau-d\tau/2)) \nonumber\\
	&=  \p_\mu f(X) \, \rd_0 X^\mu + \frac{1}{4} \, \p_\nu \p_\mu f(X) \left( \rd_+ X^\mu \rd_+X^\nu - \rd_- X^\mu \rd_-X^\nu \right) + o(d\tau) \, . 
\end{align}
\par

Using these results, one can calculate the limits. This yields
\begin{align}\label{eq:DiffFunctionsExp+-0}
	\lim_{d\tau\rightarrow 0 } \E\left[ \frac{\rd_+ f(X(\tau))}{d\tau} \, \Big| \, X(\tau)=x\right] 
	&= \left[ v_+^\mu(x,\tau) \, \p_\mu + \frac{1}{2} \, v_{2+}^{\mu\nu}(x,\tau)\, \p_\nu\p_\mu \right] f(x) \, ,
	\nonumber\\
	\lim_{d\tau\rightarrow 0 } \E\left[ \frac{\rd_- f(X(\tau))}{d\tau} \, \Big| \, X(\tau)=x\right] 
	&= \left[ v_-^\mu(x,\tau) \, \p_\mu + \frac{1}{2} \, v_{2-}^{\mu\nu}(x,\tau)\, \p_\nu\p_\mu \right] f(x) \, ,
	\nonumber\\
	\lim_{d\tau\rightarrow 0 } \E\left[ \frac{\rd_0 f(X(\tau))}{d\tau} \, \Big| \, X(\tau)=x\right] 
	&= \left[ v_\circ^\mu(x,\tau) \, \p_\mu + \frac{1}{2} \, v_{2\circ}^{\mu\nu}(x,\tau) \, \p_\nu\p_\mu \right] f(x) \, ,
\end{align}
where we defined
\begin{equation}
	v_{2\circ} := \frac{v_{2+}+v_{2-}}{2} \, .
\end{equation}
\par

The results can be generalized to differential operators $\rd_a^kf$ for any $k\in\mathbb{N}$, which is done in appendix \ref{Ap:HighOrderDiffOp}. Here, we only present the results for $k=2$:
\begin{align}
	\rd_\pm^2 f(X) &= \p_\mu f(X) \, \rd_\pm^2X^\mu + o(d\tau) \, ,\nonumber\\
	\rd_0^2 f(X) &= \p_\mu f(X) \, \rd_0^2 X^\mu 
	+ \frac{1}{2} \, \p_\nu\p_\mu f(X) \, (\rd_+X^\mu \rd_+X^\nu + \rd_-X^\mu \rd_-X^\nu) + o(d\tau) \, .
\end{align}
Then, after taking the limits one obtains
\begin{align}\label{eq:DiffFunctionsExpperp}
	\lim_{d\tau\rightarrow 0 } \E\left[ \frac{\rd_\pm^2 f(X(\tau))}{2\,d\tau} \, \Big| \, X(\tau)=x\right] 
	&=  0\, , \nonumber\\
	\lim_{d\tau\rightarrow 0 } \E\left[ \frac{\rd_0^{2} f(X(\tau))}{2\,d\tau} \, \Big| \, X(\tau)=x\right] 
	&= \left[
	v_\perp^\mu(x,\tau) \, \p_\mu 
	+ \frac{1}{2} \, v_{2\perp}^{\mu\nu}(x,\tau) \, \p_\nu\p_\mu 
	\right] f(x) \, ,
\end{align}
where we defined
\begin{equation}
	v_{2\perp} :=\frac{v_{2+}-v_{2-}}{2} \, .
\end{equation}

\section{Second order geometry}\label{eq:2ndGeom}
In section \ref{sec:Differentials}, we constructed differentials $\rd_aX$ along a `quantum trajectory' $X\in\mathcal{C}^{1/2}(\mathcal{T},\M)$ and discussed its properties. Then, in section \ref{sec:DiffOperators}, we used this to construct differential operators $\rd_a$ that act on functions $f\in\mathcal{C}^2(\M,\R)$, and evaluated $\rd_a f$ along the quantum trajectories. In this discussion we encountered various objects: $b_\pm,v_\pm,v_{2\pm}$ and linear combinations thereof, but we kept this discussion on a formal level without specifying the mathematical nature of these objects. In this section, we will partly remedy this by showing that the objects $(v,v_2)_\pm$ are generalized vectors that can be defined within the framework of second order geometry \cite{Meyer:1981,Schwartz:1984,Emery:1989,Huang:2022}. 
\par 

In a classical theory, one can define along any continuously differentiable trajectory \eqref{eq:DefDiffpm} a derivative
\begin{equation}\label{eq:ClassDer}
	\dot{X}(\tau) = \frac{\rd X(\tau)}{d\tau} \, ,
\end{equation}
such that the differentials defined in eq.~\eqref{eq:DefDiffpm} are given by\footnote{If the trajectory is only differentiable instead of continuously differentiable, one has $\rd_\pm X = \dot{X}_\pm \, d\tau + o(d\tau)$, as the derivative \eqref{eq:ClassDer} still exists but is no longer unique. If, on the other hand, the trajectory is twice differentiable, one can write $\rd_\pm X = \dot{X} \, d\tau + O(d\tau^2)$.}
\begin{equation}\label{eq:ClassDiff}
	\rd_\pm X^\mu(\tau) = \dot{X}^\mu(\tau) \, d\tau + o(d\tau) \, .
\end{equation}
Similarly, the action of the differential operators introduced in eq.~\eqref{eq:DiffOpFnct} is given by
\begin{equation}\label{eq:DiffFuncClass}
	\rd_\pm f(X) = \p_\mu f(X)\, \rd_\pm X^\mu + o(d\tau)\,,
\end{equation}
such that the total derivative along of the function $f$ is given by
\begin{equation}
	\frac{\rd f(X)}{\rd\tau} = \dot{X}^\mu \, \p_\mu f(X)\,.
\end{equation}
For any $\tau\in\mathcal{T}$, the velocity \eqref{eq:ClassDer} is an element of the tangent space $\dot{X}(\tau)\in T_{X(\tau)}\M$. The velocity along the trajectory can be generalized to a velocity field $v$, whose components are given by\footnote{As we consider a classical deterministic trajectory the expectation value may be omitted, but the conditioning is necessary for extending $\dot{X}$ to a velocity field $v(X)$.}
\begin{equation}\label{eq:VelFieldClass}
	v^\mu(x,\tau) := \E[ \dot{X}^\mu(\tau) \, |\, X(\tau) = x] \, ,
\end{equation}
which describes the velocity $\dot{X}$ at any point $x\in\M$ on the manifold. This velocity forms a section of the tangent bundle $T\M=\bigsqcup_{x\in\M}T_x\M$, and any such section $v\in\Gamma(T\M)$ defines a derivation on scalar fields $f\in \mathcal{C}^\infty(\M,\R)$ given by
\begin{equation}\label{eq:ClassAcVect}
	v : \mathcal{C}^\infty(\M,\R) \rightarrow \mathcal{C}^\infty(\M,\R)
	\qquad {\rm s.t.} \qquad
	f \mapsto v^\mu \p_\mu f \, .
\end{equation}
Furthermore, using these vector fields, one can define the differential operators $\rd$ 
\begin{equation}\label{eq:DiffOpFO}
	\rd: \mathcal{C}^\infty(\M,\R) \rightarrow T^\ast\M 
	\qquad {\rm s.t.} \qquad
	f \mapsto \rd f = \p_\mu f \, \rd x^\mu \, ,
\end{equation}
where $T^\ast\M$ is the cotangent bundle, which is dual to $T\M$.
\par 

Using the results from previous section, one can now generalize this discussion to trajectories that are $\alpha$-H\"older continuous for $\alpha<1/2$. In this case, the classical differential \eqref{eq:ClassDiff} is replaced by eq.~\eqref{eq:DifferentialImposed}, whereas the differential \eqref{eq:DiffFuncClass} must be replaced by eqs.~\eqref{eq:DiffFunc+} and~\eqref{eq:DiffFunc-}. Moreover, using eqs.~\eqref{eq:DiffFunctionsExp+-0} and \eqref{eq:DiffFunctionsExpperp}, one finds a natural generalization of the classical action of vector fields on scalar functions eq.~\eqref{eq:ClassAcVect} to 
\begin{equation}
	(v,v_2) : \mathcal{C}^\infty(\M,\R) \rightarrow \mathcal{C}^\infty(\M,\R)
	\qquad {\rm s.t.} \qquad
	f \mapsto \left( v^\mu + \frac{1}{2} v_2^{\mu\nu} \p_\nu \right) \p_\mu f \, .
\end{equation}
This action no longer defines a derivation, as it violates Leibniz rule, but these vector fields $(v,v_2)\in\Gamma(T_2\M)$ can be regarded as sections of a generalized vector bundle, called the \textit{second order vector bundle} $T_2\M$. 
\par 

It is natural to also generalize the differential operator \eqref{eq:DiffOpFO} to a second order differential operator given by
\begin{equation}\label{eq:DiffOpSecOrder}
	\rd_2: \mathcal{C}^\infty(\M,\R) \rightarrow T_2^\ast\M 
	\qquad {\rm s.t.} \qquad
	f \mapsto \rd_2f = \p_\mu f \, \rd_2 x^\mu + \frac{1}{2} \, \p_\nu \p_\mu f \, \rd x^\mu \rd x^\nu \, ,
\end{equation}
where $T_2^\ast\M$ is the second order cotangent bundle that is dual to $T_2\M$.
Instead of satisfying the Leibniz rule, this operator satisfies a generalized product rule of the form
\begin{equation}
	\rd_2(f\,g) = f\, \rd_2g + \rd_2f \, g + \rd f\, \rd g \, ,
\end{equation}
where $\rd$ is the first order differential \eqref{eq:DiffOpFO}. Similarly, for functions $h\in C^\infty(\R,\R)$ one obtains a generalized chain rule of the form
\begin{equation}
	\rd_2(h \circ f) = (h'\circ f)\, \rd_2f + \frac{1}{2} \, (h''\circ f) \, \rd f \, \rd f\, .
\end{equation}
\par 

We remark that we introduced second order vectors $(v,v_2)$, whereas in previous section we found the objects $(v_+,v_{2+})$, $(v_-,v_{2-})$ and $(v_\circ,v_{2\circ})$, $(v_\perp,v_{2\perp})$. Clearly, all these objects are second order vectors, and the above discussion applies to all of them.

\subsection{Second order (co)tangent spaces}
As in the first order case, the second order (co)tagent bundle can be regarded as a bundle of second order (co)tangent spaces:
\begin{equation}\label{eq:SecondorderTBundle}
	T_2\M = \bigsqcup_{x\in\M} T_{2,x}\M 
	\qquad {\rm and} \qquad
	T_2^\ast\M = \bigsqcup_{x\in\M} T^\ast_{2,x}\M \, .
\end{equation}
We will now construct bases for these spaces, and discuss the duality pairing between those bases. We first note that the dimension of the manifold and first order spaces is given by
\begin{equation}
	{\rm dim}(T_x^\ast\M) = {\rm dim}(T_x\M) = {\rm dim}(\M) = n \, .
\end{equation}
The second order tangent spaces, on the other hand, have extra degrees of freedom encoded in the second order part of the vector $v_2$. Then, using the commutativity of partial derivatives, one finds that these second order parts are symmetric, such that the dimension of the (co)tangent spaces is given by
\begin{equation}\label{eq:dimTangentsecond}
	{\rm dim}(T_{2,x}\M) = {\rm dim}(T^\ast_{2,x}\M) = n + \frac{n(n+1)}{2} = \frac{n(n+3)}{2} =: N \, .
\end{equation}
On the tangent space we may construct the following coordinate basis
\begin{equation}\label{eq:BasisTangent}
	\Big\{\p_\mu,\p_{\mu\mu}, \frac{\p_{\mu\nu}+ \p_{\nu\mu}}{\sqrt{2}} \, \Big| \, \mu<\nu \in\{0,...,n-1\}\Big\} \, ,
\end{equation}
such that vectors $v\in T_{2,x}\M$ are expanded with respect to this basis as
\begin{align}\label{eq:VectorExp}
	v 
	&= 
	\sum_{\mu=0}^d v^\mu \, \p_\mu 
	+ \sum_{\mu=0}^d v_2^{\mu\mu} \, \p_{\mu\mu}
	+ \sum_{\mu=0}^d \sum_{\nu=\mu+1}^d  \sqrt{2} \, v_2^{\mu\nu} \frac{ \p_{\mu\nu} + \p_{\nu\mu}}{\sqrt{2}} \nonumber\\
	&= v^\mu \, \p_\mu + v_2^{\mu\nu} \, \p_{\mu\nu} \, ,
\end{align}
where Einstein's summation convention is applied in the second line, and where we imposed the symmetry $v_2^{\mu\nu} = v_2^{\nu\mu}$.
\par

Similarly, on the the cotangent space, we construct a coordinate basis\footnote{In the literature various notations exist for the second order basis vectors. E.g. in Ref.~\cite{Emery:1989} one finds $\rd x^\mu \cdot \rd x^\nu$ and in Ref.~\cite{Kuipers:2023pzm}, one finds $\rd[x^\mu,x^\nu]$ instead of $\rd\langle x^\mu,x^\nu\rangle$.}
\begin{equation}\label{eq:eq:BasisCoTangent}
	\Big\{\rd_2x^\mu,\rd\langle x^\mu,x^\mu\rangle, \frac{\rd\langle x^\mu,x^\nu\rangle + \rd\langle x^\nu,x^\mu\rangle}{\sqrt{2}} \, \Big| \, \mu<\nu \in\{0,...,n-1\}\Big\} \,,
\end{equation}
such that covectors $\omega\in T_{2,x}^\ast\M$ are expanded with respect to this basis as
\begin{align}\label{eq:FormExp}
	\omega &= 
	\sum_{\mu=0}^d \omega_\mu \, \rd_2x^\mu 
	+ \sum_{\mu=0}^d \omega_{\mu\mu} \, \rd\langle x^\mu, x^\mu \rangle 
	+ \sum_{\mu=0}^d \sum_{\nu=\mu+1}^d \sqrt{2}\, \omega_{\mu\nu} \frac{ \rd\langle x^\mu, x^\nu \rangle  + \rd\langle x^\nu, x^\mu \rangle }{\sqrt{2}}  \nonumber\\
	&= \omega_\mu \, \rd_2x^\mu + \omega_{\mu\nu} \, \rd\langle x^\mu, x^\nu \rangle \, ,
\end{align}
where Einstein's summation convention is applied in the second line and where we imposed the symmetry $\omega_{\mu\nu}=\omega_{\nu\mu}$.
\par 

We define a duality pairing between the tangent and cotangent space by 
\begin{equation}\label{eq:Dualitypairing}
	\begin{tabular}{ c | c c }
		$\langle .,. \rangle$ & $ \rd_2 x^\nu$ & $\rd\langle x^\nu,x^\sigma \rangle$ \\
		\hline
		$\p_\mu $ & $\delta_\mu^\nu$ & $0$ \\ 
		$\p_{\mu\rho}$ & $0$ & $\delta_\mu^\nu \delta_\rho^\sigma $   
	\end{tabular}
\end{equation}
such that the pairing of the basis elements is given by\footnote{Note that $\mu<\rho$ and $\nu<\sigma$, such that $\delta_\mu^\sigma\delta_\nu^\rho=0$.}
\begin{equation}
	\begin{tabular}{ c | c c c}
		$\langle .,. \rangle$ & $ \rd_2 x^\nu$ & $\rd\langle x^\nu,x^\nu \rangle$ & $\frac{\rd\langle x^\nu,x^\sigma \rangle + \rd\langle x^\sigma , x^\nu \rangle}{\sqrt{2}}$ \\
		\hline
		$\p_\mu $ & $\delta_\mu^\nu$ & $0$ & $0$\\ 
		$\p_{\mu\mu}$ & $0$ & $\delta_\mu^\nu$ & $0$  \\
		$\frac{\p_{\mu\rho}+\p_{\rho\mu}}{\sqrt{2}}$ & $0$ & $0$ & $\delta_\mu^\nu \delta_\rho^\sigma  $  
	\end{tabular}
\end{equation}
and a generic pairing between vectors and forms by
\begin{equation}\label{eq:Pairing}
	\langle \omega, \, v \rangle = \omega_\mu v^\mu +  \omega_{\mu\nu} v_2^{\mu\nu}\, .
\end{equation}

\subsection{Isomorphisms with first order geometry}
The second order vectors contain two parts: a first order part that looks like a regular first order vector and a second order part that looks like a symmetric $(2,0)$-tensor. Similarly, covectors contain a first order part that looks like a first order covector and a second order part that looks like a $(0,2)$-tensor. This observation can be made precise by the following isomorphisms  \cite{Huang:2022}
\begin{align}\label{eq:TangentSpaceDecomp}
	T_{2,x}\M &\cong T_x\M \oplus {\rm Sym}(T_x\M \otimes T_x\M) \, ,\nonumber\\
	T^\ast_{2,x}\M &\cong T^\ast_x\M \oplus {\rm Sym}(T^\ast_x\M \otimes T^\ast_x\M)\,,
\end{align}
which only holds fiber-wise on the tangent bundle $T_2\M=\bigsqcup_{x\in\M}T_{2,x}\M$, since the structure group on this bundle is the It\^o group that couples the two sectors.
\par 

Due to the existence of these isomorphisms, we can define various mappings between first and second order geometry \cite{Emery:1989}. We first note that second order vectors can be obtained as products of first vectors: given two vectors $u,v\in T_x\M$ their product is given by
\begin{equation}
	u \, v = (u^\nu \p_\nu v^\mu) \, \p_\mu + 2 \,(u^\nu v^\mu) \, \p_{\mu\nu} \, .
\end{equation}
Moreover, second order forms can be projected onto first order forms by:
\begin{equation}
	\mathcal{P} : T_{2,x}^\ast\M \rightarrow T_x^\ast\M
	\qquad {\rm s.t.} \qquad
	\omega_\mu \, \rd_2x^\mu + \omega_{\mu\nu} \, \rd\langle x^\mu, x^\nu \rangle
	\mapsto
	\omega_\mu \rd x^\mu \, .
\end{equation}
One can also define a type of differential operator that maps first order forms into second order forms:
\begin{equation}
	\underline{\rd} : T_x^\ast\M \rightarrow T_{2,x}^\ast\M
	\qquad {\rm s.t.} \qquad
	\omega_\mu  \rd x^\mu \mapsto \omega_\mu \rd_2x^\mu + \frac{1}{2} \, \p_\nu \omega_\mu \, \rd\langle x^\mu, x^\nu \rangle \, .
\end{equation}
This may be composed with the usual exterior derivative, which yields
\begin{equation}\label{eq:d2f}
	\rd_2 f = \underline{\rd} (\rd f) = \p_\mu f \, \rd_2x^\mu + \frac{1}{2} \, \p_\nu\p_\mu f \,  \rd\langle x^\mu, x^\nu \rangle\,.
\end{equation}
In addition, one can construct a map involving the second order parts of the vectors and forms, given by
\begin{equation}\label{eq:MapH}
	\mathcal{H}: {\rm Sym}(T_x^\ast\M \otimes T_x^\ast\M) \rightarrow T_{2,x}^\ast\M
	\qquad {\rm s.t. } \qquad
	\omega_{\mu\nu}\, \rd x^\mu \otimes \rd x^\nu \mapsto \omega_{\mu\nu} \, \rd\langle x^\mu, x^\nu \rangle \, .
\end{equation}
By taking the dual of this map, one obtains a projection of second order vectors on (2,0)-tensors:
\begin{equation}
	\mathcal{H}^\ast:  T_{2,x}\M \rightarrow {\rm Sym}(T_x\M \otimes T_x\M)
	\qquad {\rm s.t. } \qquad
	v^\mu \p_\mu +  v_2^{\mu\nu} \, \p_{\mu\nu} \mapsto \frac{1}{2} \, v_2^{\mu\nu} \, \p_\mu \otimes \p_\nu \, .
\end{equation}
Using these maps, we can formally equate\footnote{In the literature, one finds various other conventions that differ by a factor $2$, cf. e.g. \cite{Emery:1989,Kuipers:2023pzm,Bies:2023zvs}. These conventions affect many of the expressions discussed in this section \ref{eq:2ndGeom}. This work aligns with the conventions from Ref.~\cite{Bies:2023zvs}. These differ from Refs.~\cite{Emery:1989,Kuipers:2023pzm}, where $\p_{\mu\nu}=\p_\nu\p_\mu$ and the factor $1/2$ is included in different places: in Ref.~\cite{Emery:1989} it is included in the cotangent basis elements such that $\rd\langle x,x\rangle = \frac{dx^\mu dx^\nu}{2}$, while in Ref.~\cite{Kuipers:2023pzm} it is explicitly included in the expansion of vectors, such that $v=v^\mu \p_\mu + \frac{1}{2} v_2^{\mu\nu} \p_{\mu\nu}$. }
\begin{align}\label{Informalpmupnu}
	\p_{\mu\nu} &= \frac{\p_\nu\p_\mu}{2} \, ,\nonumber\\
	\rd \langle x^\mu, x^\nu\rangle &= \rd x^\mu \rd x^\nu.
\end{align}

\subsection{Covariance}

In this section we will study the behaviour of second order vectors and forms under a general coordinate transformation $x^\mu\rightarrow\tilde{x}^\mu$.

\subsubsection{Vectors} 
For a generic second order vector, one finds
\begin{align}
	v 
	&= v^\mu \p_\mu + v_2^{\mu\nu} \p_{\mu\nu} \nonumber\\
	&=
	v^\mu \frac{\p \tilde{x}^\nu}{\p x^\mu} \tilde{\p}_\nu 
	+ \frac{1}{2} \, v_2^{\mu\nu} \p_\nu \left( \frac{\p \tilde{x}^\rho}{\p x^\mu} \tilde{\p}_\rho \right) \nonumber\\
	&= \left( 
		v^\nu \frac{\p \tilde{x}^\mu}{\p x^\nu} 
		+ \frac{1}{2} \, v_2^{\nu\rho} \frac{\p^2 \tilde{x}^\mu}{\p x^\nu \p x^\rho}
	\right) \tilde{\p}_\mu
	+ v_2^{\rho\sigma} \frac{\p \tilde{x}^\nu}{\p x^\sigma} \frac{\p \tilde{x}^\mu}{\p x^\rho} \, \tilde{\p}_{\mu\nu} \, .
\end{align}
Hence, vectors transform as
\begin{equation}\label{eq:Vec2Trans}
	\begin{pmatrix}
		v^\mu \\ v_2^{\rho\sigma}
	\end{pmatrix}
	\rightarrow 
	\begin{pmatrix}
		\tilde{v}^\mu \\ \tilde{v}_2^{\rho\sigma}
	\end{pmatrix}
	=
	\begin{pmatrix}
		\frac{\p \tilde{x}^\mu}{\p x^\nu} 
		& \frac{1}{2} \frac{\p^2 \tilde{x}^\mu}{\p x^\kappa\p x^\lambda} \\
		0 &
		\frac{\p \tilde{x}^\rho}{\p x^\kappa} \frac{\p \tilde{x}^\sigma}{\p x^\lambda}
	\end{pmatrix}
	\begin{pmatrix}
		v^\nu \\ v_2^{\kappa\lambda}
	\end{pmatrix},
\end{equation}
whereas the basis elements transform as
\begin{equation}
	\begin{pmatrix}
		\p_\mu \\ \p_{\rho\sigma}
	\end{pmatrix}
	\rightarrow 
	\begin{pmatrix}
		\tilde{\p}_\mu \\ \tilde{\p}_{\rho\sigma}
	\end{pmatrix}
	=
	\begin{pmatrix}
		\frac{\p x^\nu}{\p \tilde{x}^\mu} 
		& 0 \\
		\frac{1}{2} \frac{\p^2 x^\nu}{\p \tilde{x}^\rho\p \tilde{x}^\sigma} &
		\frac{\p x^\kappa}{\p \tilde{x}^\rho} \frac{\p x^\lambda}{\p x^\sigma}
	\end{pmatrix}
	\begin{pmatrix}
		\p_\nu \\ \p_{\kappa\lambda}
	\end{pmatrix}.
\end{equation}
\par 

The fact that the transformation matrices are not diagonal implies that the first order part $v^\mu$ and second order part $\p_{\mu\nu}$ do not transform contravariantly. However, this can easily be remedied by introducing a contravariant vector representation:
\begin{alignat}{3}\label{eq:CovVec}
	\hat{v}^\mu &:= v^\mu + \frac{1}{2} \, \Gamma^\mu_{\rho\sigma} v_2^{\rho\sigma} \, ,
	\qquad \qquad && \hat{\p}_\mu &&:= \p_\mu \, , \nonumber\\
	\hat{v}_2^{\mu\nu} &:= v_2^{\mu\nu} \, ,
	\qquad \qquad && \hat{\p}_{\mu\nu} &&:= \p_{\mu\nu} - \frac{1}{2} \, \Gamma^\rho_{\mu\nu} \p_\rho  \, ,
\end{alignat}
where $\Gamma$ is the Christoffel symbol describing the affine connection on the first order tangent bundle.
Then,
\begin{align}
	v 
	&= v^\mu \p_\mu + v_2^{\mu\nu} \p_{\mu\nu} \nonumber\\
	&= \hat{v}^\mu \hat{\p}_\mu + \hat{v}_2^{\mu\nu} \hat{\p}_{\mu\nu} 
\end{align}
with covariant transformation laws
\begin{equation}
	\begin{pmatrix}
		\hat{v}^\mu \\ \hat{v}_2^{\rho\sigma}
	\end{pmatrix}
	\rightarrow 
	\begin{pmatrix}
		\hat{\tilde{v}}^\mu \\ \hat{\tilde{v}}_2^{\rho\sigma}
	\end{pmatrix}
	=
	\begin{pmatrix}
		\frac{\p \tilde{x}^\mu}{\p x^\nu} 
		& 0 \\
		0 &
		\frac{\p \tilde{x}^\rho}{\p x^\kappa} \frac{\p \tilde{x}^\sigma}{\p x^\lambda}
	\end{pmatrix}
	\begin{pmatrix}
		\hat{v}^\nu \\ \hat{v}_2^{\kappa\lambda}
	\end{pmatrix}
\end{equation}
and
\begin{equation}
	\begin{pmatrix}
		\hat{\p}_\mu \\ \hat{\p}_{\rho\sigma}
	\end{pmatrix}
	\rightarrow 
	\begin{pmatrix}
		\hat{\tilde{\p}}_\mu \\ \hat{\tilde{\p}}_{\rho\sigma}
	\end{pmatrix}
	=
	\begin{pmatrix}
		\frac{\p x^\nu}{\p \tilde{x}^\mu} 
		& 0 \\
		0 &
		\frac{\p x^\kappa}{\p \tilde{x}^\rho} \frac{\p x^\lambda}{\p x^\sigma}
	\end{pmatrix}
	\begin{pmatrix}
		\hat{\p}_\nu \\ \hat{\p}_{\kappa\lambda}
	\end{pmatrix}.
\end{equation}

\subsubsection{Forms}
For a generic second order form, one can study its transformation using the mapping $\underline{\rd}$. One finds
\begin{align}
	\underline{\rd} \left(\omega_\mu \, \rd x^\mu\right)
	&\rightarrow
	\underline{\rd}\left( \tilde{\omega}_\nu \, \frac{\p \tilde{x}^\nu}{\p x^\mu} \, \rd x^\mu\right)
	\nonumber\\
	&=
	\tilde{\omega}_\nu \frac{\p \tilde{x}^\nu}{\p x^\mu} \, \rd_2 x^\mu
	+ \frac{1}{2} \, \frac{\p}{\p x^\rho} \left( 
	\tilde{\omega}_\nu \, \frac{\p \tilde{x}^\nu}{\p x^\mu}
	\right) \rd\langle x^\mu , x^\rho \rangle
	\nonumber\\
	&=
	\left( \tilde{\omega}_\nu \frac{\p \tilde{x}^\nu}{\p x^\mu} \right) \rd_2 x^\mu
	+ \frac{1}{2} \left( \tilde{\omega}_\rho \frac{\p^2 \tilde{x}^\rho}{\p x^\nu \p x^\mu} 
	+ \tilde{\omega}_{\rho\sigma} 
	\frac{\p \tilde{x}^\rho}{\p x^\mu} \frac{\p \tilde{x}^\sigma}{\p x^\nu}
	\right) \rd\langle x^\mu , x^\nu \rangle \, .
\end{align}
Hence, forms transform as
\begin{equation}
	\begin{pmatrix}
		\omega_\mu \\ \omega_{\rho\sigma}
	\end{pmatrix}
	\rightarrow 
	\begin{pmatrix}
		\tilde{\omega}_\mu \\ \tilde{\omega}_{\rho\sigma}
	\end{pmatrix}
	=
	\begin{pmatrix}
		\frac{\p x^\nu}{\p \tilde{x}^\mu} 
		& 0 \\
		\frac{1}{2} \frac{\p^2 x^\nu}{\p \tilde{x}^\rho\p \tilde{x}^\sigma} &
		\frac{\p x^\kappa}{\p \tilde{x}^\rho} \frac{\p x^\lambda}{\p x^\sigma}
	\end{pmatrix}
	\begin{pmatrix}
		\omega_\nu \\ \omega_{\kappa\lambda}
	\end{pmatrix},
\end{equation}
whereas the basis elements transform as
\begin{equation}
	\begin{pmatrix}
		\rd_2x^\mu \\ \rd\langle x^\rho,x^\sigma\rangle
	\end{pmatrix}
	\rightarrow 
	\begin{pmatrix}
		\rd_2\tilde{x}^\mu \\ \rd\langle\tilde{x}^\rho,\tilde{x}^\sigma\rangle
	\end{pmatrix}
	=
	\begin{pmatrix}
		\frac{\p \tilde{x}^\mu}{\p x^\nu} 
		& \frac{1}{2} \frac{\p^2 \tilde{x}^\mu}{\p x^\kappa\p x^\lambda} \\
		0 &
		\frac{\p \tilde{x}^\rho}{\p x^\kappa} \frac{\p \tilde{x}^\sigma}{\p x^\lambda}
	\end{pmatrix}
	\begin{pmatrix}
		\rd_2x^\nu \\ \rd\langle x^\kappa,x^\lambda\rangle
	\end{pmatrix}.
\end{equation}
\par 

The fact that the transformation matrices are not diagonal implies that the first order part $\rd_2x^\mu$ and second order part $\omega_{\mu\nu}$ do not transform covariantly. However, this can easily be remedied by introducing a covariant form representation:
\begin{alignat}{3}\label{eq:CovForm}
	\hat{\omega}_\mu &:= \omega_\mu \, ,
	\qquad \qquad && \rd_2\hat{x}^\mu &&:= \rd_2x^\mu + \frac{1}{2} \, \Gamma^\mu_{\rho\sigma} \, \rd\langle x^\rho,x^\sigma \rangle \, ,\nonumber\\
	\hat{\omega}_{\mu\nu} &:= \omega_{\mu\nu} - \frac{1}{2} \Gamma^\rho_{\mu\nu} \omega_\rho \, ,
	\qquad \qquad && \rd\langle \hat{x}^\mu,\hat{x}^\nu\rangle &&:= \rd\langle x^\mu,x^\nu\rangle  \, .
\end{alignat}
Then,
\begin{align}
	\omega 
	&= \omega_\mu \, \rd_2x^\mu + \omega_{\mu\nu} \, \rd\langle x^\mu,x^\nu\rangle  \nonumber\\
	&= \hat{\omega}_\mu \, \rd_2\hat{x}^\mu + \hat{\omega}_{\mu\nu} \, \rd\langle \hat{x}^\mu,\hat{x}^\nu\rangle 
\end{align}
with covariant transformation laws
\begin{equation}
	\begin{pmatrix}
		\hat{\omega}_\mu \\ \hat{\omega}_{\rho\sigma}
	\end{pmatrix}
	\rightarrow 
	\begin{pmatrix}
		\hat{\tilde{\omega}}_\mu \\ \hat{\tilde{\omega}}_{\rho\sigma}
	\end{pmatrix}
	=
	\begin{pmatrix}
		\frac{\p x^\nu}{\p \tilde{x}^\mu} 
		& 0 \\
		0 &
		\frac{\p x^\kappa}{\p \tilde{x}^\rho} \frac{\p x^\lambda}{\p x^\sigma}
	\end{pmatrix}
	\begin{pmatrix}
		\hat{\omega}_\nu \\ \hat{\omega}_{\kappa\lambda}
	\end{pmatrix}
\end{equation}
and
\begin{equation}
	\begin{pmatrix}
		\rd_2\hat{x}^\mu \\ \rd \langle\hat{x}^\rho,\hat{x}^\sigma \rangle
	\end{pmatrix}
	\rightarrow 
	\begin{pmatrix}
		\rd_2\hat{\tilde{x}}^\mu \\ \rd\langle \hat{\tilde{x}}^\rho,\hat{\tilde{x}}^\sigma \rangle
	\end{pmatrix}
	=
	\begin{pmatrix}
		\frac{\p \tilde{x}^\mu}{\p x^\nu} 
		& 0 \\
		0 &
		\frac{\p \tilde{x}^\rho}{\p x^\kappa} \frac{\p \tilde{x}^\sigma}{\p x^\lambda}
	\end{pmatrix}
	\begin{pmatrix}
		\rd_2\hat{x}^\nu \\ \rd\langle \hat{x}^\kappa,\hat{x}^\lambda\rangle
	\end{pmatrix} .
\end{equation}
\subsubsection{Covariant mappings}

Given the transformation laws discussed in the previous two subsections, we can define a covariant projection of second order vectors onto first order vectors \cite{Emery:1989}, given by
\begin{equation}
	\mathcal{F} : T_{2,x}\M \rightarrow T_x\M 
	\qquad {\rm s.t.} \qquad 
	v^\mu \p_\mu + v_2^{\mu\nu} \p_{\mu\nu} \mapsto \left(v^\mu + \frac{1}{2} \, \Gamma^\mu_{\nu\rho} v_2^{\nu\rho}\right) \p_\mu \, .
\end{equation}
By taking the dual, one obtains a map from first order forms to second order forms: 
\begin{equation}
	\mathcal{F}^\ast : T_x^\ast\M \rightarrow  T_{2,x}^\ast\M
	\qquad {\rm s.t.} \qquad 
	\omega_\mu \rd x^\mu \mapsto \omega_\mu \, \rd_2 x^\mu + \frac{1}{2} \, \omega_\rho \Gamma^\rho_{\mu\nu} \, \rd\langle x^\mu,x^\nu\rangle   \, .
\end{equation}

Furthermore, since we have constructed new basis elements in eqs.~\eqref{eq:CovVec} and \eqref{eq:CovForm}, we may also consider their pairing which is given by
\begin{equation}
	\begin{tabular}{ c | c c }
		$\langle .,. \rangle$ & $ \rd_2 \hat{x}^\nu$ & $\rd\langle \hat{x}^\nu,\hat{x}^\sigma \rangle$ \\
		\hline
		$\hat{\p}_\mu $ & $\delta_\mu^\nu$ & $0$ \\ 
		$\hat{\p}_{\mu\rho}$ & $0$ & $\delta_\mu^\nu \delta_\rho^\sigma $   
	\end{tabular}
\end{equation}
Hence, a generic pairing of vectors and forms is given by
\begin{equation}
	\langle \omega, \, v \rangle = \hat{\omega}_\mu \hat{v}^\mu + \hat{\omega}_{\mu\nu} \hat{v}^{\mu\nu}\, ,
\end{equation}
which is consistent with eq.~\eqref{eq:Pairing}.

\section{Second order metric}\label{sec:Metric}
In this section, we will construct an inner product on the tangent spaces $T_{2,x}\M$. This inner product defines a second order metric $G$ and acts as
\begin{equation}
	\begin{tabular}{ c | c c }
		$\cdot$ & $ \p_\nu$ & $\p_{\kappa\lambda}$ \\
		\hline
		$\p_\mu$ & $G_{\mu\nu}$ 
		& $G_{\mu|\kappa\lambda|}$ 
		\\ 
		$\p_{\rho\sigma}$ 
		& $G_{|\rho\sigma|\nu} $ 
		& $G_{|\rho\sigma|\kappa\lambda|} $   
	\end{tabular}
\end{equation}
such that the inner product of two second vectors is given by
\begin{align}\label{eq:MetricInProd}
	G(u,v) 
	&= G_{\mu\nu} u^\mu v^\nu 
	+ G_{\mu|\kappa\lambda|} u^\mu v_2^{\kappa\lambda} 
	+ G_{|\rho\sigma|\nu} u_2^{\rho\sigma} v^\nu 
	+ G_{|\rho\sigma|\kappa\lambda|} u_2^{\rho\sigma} v_2^{\kappa\lambda}\, ,
\end{align}
Similarly, on the cotangent space, we define
\begin{equation}
	\begin{tabular}{ c | c c }
		$\cdot$ & $ \rd_2 x^\nu$ & $\rd\langle \hat x^\kappa,x^\lambda \rangle$ \\
		\hline
		$\rd_2x^\mu$ 
		& $G^{\mu\nu}$ 
		& $G^{\mu|\kappa\lambda|}$
		\\ 
		$\rd\langle x^\rho , x^\sigma\rangle$ 
		& $G^{|\rho\sigma|\nu}$ 
		& $G^{|\rho\sigma|\kappa\lambda|}$   
	\end{tabular}
\end{equation}
such that the inner product of two forms is given by
\begin{align}
	G(\alpha,\beta) 
	&= G^{\mu\nu} \alpha_\mu \beta_\nu 
	+ G^{\mu|\kappa\lambda|} \alpha_\mu \beta_{\kappa\lambda} 
	+ G^{|\rho\sigma|\nu} \alpha_{\rho\sigma} \beta_\nu 
	+ G^{|\rho\sigma|\kappa\lambda|} \alpha_{\rho\sigma} \beta_{\kappa\lambda} \, .
\end{align}
On the basis \eqref{eq:BasisTangent} the metric acts as\footnote{Note that $\mu<\rho$ and $\nu<\sigma$ in this expression, and that the brackets $()$ denote symmetrization.}
\begin{equation}\label{eq:MetricDef}
	\begin{tabular}{ c | c c c}
		$\cdot$ & $ \p_\nu$ & $\p_{\nu\nu}$ & $\frac{\p_{\nu\sigma} + \p_{\sigma\nu}}{\sqrt{2}}$ \\
		\hline
		$\p_\mu $ & $G_{\mu\nu}$ & $G_{\mu(\nu\nu)}$ & $\sqrt{2} \, G_{\mu(\nu\sigma)}$\\ 
		$\p_{\mu\mu}$ & $G_{(\mu\mu)\nu}$ & $G_{(\mu\mu)(\nu\nu)}$ & $ \sqrt{2} \, G_{(\mu\mu)(\nu\sigma)}$  \\
		$\frac{\p_{\mu\rho}+\p_{\rho\mu}}{\sqrt{2}}$ & $\sqrt{2} \, G_{(\mu\rho)\nu}$ & $\sqrt{2} \, G_{(\mu\rho)(\nu\nu)}$ & $2 \, G_{(\mu\rho)(\nu\sigma)}$
	\end{tabular}
\end{equation}
and on the basis of the cotangent space \eqref{eq:eq:BasisCoTangent} one obtains
\begin{equation}\label{eq:CoMetricDef}
	\begin{tabular}{ c | c c c}
		$\cdot$ & $ \rd_2x^\nu$ & $\rd\langle x^\nu,x^\nu\rangle$ & $\frac{\rd\langle x^\nu,x^\sigma\rangle + \rd\langle x^\sigma,x^\nu\rangle}{\sqrt{2}}$ \\
		\hline
		$\rd_2x^\mu $ & $G^{\mu\nu}$ & $G^{\mu(\nu\nu)}$ & $\sqrt{2} \, G^{\mu(\nu\sigma)}$\\ 
		$\rd\langle x^\mu,x^\mu\rangle$ & $G^{(\mu\mu)\nu}$ & $G^{(\mu\mu)(\nu\nu)}$ & $\sqrt{2} \, G^{(\mu\mu)(\nu\sigma)}$  \\
		$\frac{\rd\langle x^\mu,x^\rho\rangle+\rd\langle x^\rho,x^\mu\rangle}{\sqrt{2}}$ & $\sqrt{2} \, G^{(\mu\rho)\nu}$ & $\sqrt{2} \, G^{(\mu\rho)(\nu\nu)}$ & $2 \, G^{(\mu\rho)(\nu\sigma)}$
	\end{tabular}
\end{equation}
\par

By construction, the metric components carry different physical dimensions, and the relation between the dimensions is given by
\begin{equation}
	[G_{|\mu\rho|\nu\sigma|}] 
	= [G_{\mu|\nu\sigma|}]\, L^{-1}
	= [G_{|\mu\rho|\nu}]\, L^{-1}
	= [G_{\mu\nu}] \, L^{-2}
	\, .
\end{equation}
Similarly, the dimension of the inverse components are related by 
\begin{equation}
	[G^{|\mu\rho|\nu\sigma|}] 
	= [G^{\mu|\nu\sigma|}]\, L
	= [G^{|\mu\rho|\nu}]\, L
	= [G^{\mu\nu}] \, L^{2}
	\, .
\end{equation}
\par 

Furthermore, the components of $G^{|^\mu_{\rho\sigma}|^\nu_{\kappa\lambda}|}$ and $G_{|^\mu_{\rho\sigma}|^\nu_{\kappa\lambda}|}$ are subjected to various constraints. 
A first constraint is that the cometric, defined on the cotangent space, is the inverse of the metric, defined on the tangent space. This constraint fixes the components of the cometric in terms of the components of the metric by the relation
\begin{equation}\label{eq:Invert}
	\begin{pmatrix}
		G_{\mu\alpha} & G_{\mu|\beta\gamma|} \\
		G_{|\rho\sigma|\alpha} & G_{|\rho\sigma|\beta\gamma|}
	\end{pmatrix}
	\begin{pmatrix}
		G^{\alpha\nu} & G^{\alpha|\kappa\lambda|} \\
		G^{|\beta\gamma|\nu} & G^{|\beta\gamma|\kappa\lambda|}
	\end{pmatrix}
	=
	\begin{pmatrix}
		\delta_\mu^\nu & 0 \\
		0 & \delta_\rho^\kappa \delta_\sigma^\lambda  
	\end{pmatrix}\,.
\end{equation}
A second constraint follows from the fact that the metric is a symmetric object, which imposes
\begin{align}\label{eq:MetricSym1}
	G_{\mu\nu} &= G_{\nu\mu} \, ,\nonumber\\
	G_{|\rho\sigma|\nu} &= G_{\nu|\rho\sigma|} \, ,\nonumber\\
	G_{|\rho\sigma|\kappa\lambda|} &= G_{|\kappa\lambda|\rho\sigma|} \, .
\end{align}
\par 

The remaining degrees of freedom of $G_{|^\mu_{\rho\sigma}|^\nu_{\kappa\lambda}|}$ can be fixed freely. However, due to the symmetry of second order vectors and forms, only the symmetrized components $G_{(^\mu_{\rho\sigma})(^\nu_{\kappa\lambda})}$ will be of relevance, as can be seen in eqs.~\eqref{eq:MetricDef} and \eqref{eq:CoMetricDef}. Moreover, in this work, we advocate a minimal extension from first order to second order geometry, in which all second order data can be expressed in terms of the first order data. For the metric function, this implies that the second order metric can be written as a function of the first order (co)metric and derivatives thereof.

\subsection{Flat space(time)}
In this subsection, we determine the metric in the flat Euclidean space $\R^n$. In this case the symmetry relation \eqref{eq:MetricSym1} together with the condition that the metric is composed of the first order (co)metric and its derivatives imposes that the metric components are of the form
\begin{equation}
	\begin{pmatrix}
		G_{ij} & G_{i|mn|} \\
		G_{|kl|j} & G_{|kl|mn|}
	\end{pmatrix}
	= 
	\begin{pmatrix}
		a_1 \, \delta_{ij} & 0 \\
		0 & l_s^{-2} \left( a_2 \, \delta_{km} \delta_{ln} + a_3 \, \delta_{kn} \delta_{lm} + a_4 \, \delta_{kl} \delta_{mn} \right)
	\end{pmatrix} \, .
\end{equation}
Similarly, the cometric is given by
\begin{equation}
	\begin{pmatrix}
		G^{ij} & G^{i|mn|} \\
		G^{|kl|j} & G^{|kl|mn|}
	\end{pmatrix}
	= 
	\begin{pmatrix}
		a_5 \, \delta^{ij} & 0 \\
		0 & l_s^2 \left( a_6 \, \delta^{km} \delta^{ln} + a_7 \, \delta^{kn} \delta^{lm} + a_8 \, \delta^{kl} \delta^{mn} \right)
	\end{pmatrix} \, ,
\end{equation}
where $a_i$ are numerical constants and $l_s$ is a length scale. We may now evaluate the invertability condition \eqref{eq:Invert}, which imposes
\begin{align}
	a_1 \, a_5 &= 1 \, ,\nonumber\\
	a_2 \, a_6 + a_3 \, a_7 &= 1 \, ,\nonumber\\
	a_2 \, a_7 + a_3 \, a_6 &= 0 \, ,\nonumber\\
	a_2 \, a_8 + a_3 \, a_8 + a_4 \, a_6 + a_4 \, a_7 + n \, a_4\, a_8 &= 0 \,. 
\end{align}
In order to fix the remaining freedom, we evaluate the inner product on the basis \eqref{eq:BasisTangent}, as given in eq.~\eqref{eq:MetricDef}, which yields\footnote{Note that $i<k$ and $j<l$ in this expression}
\begin{equation}
	\begin{tabular}{ c | c c c}
		$\cdot$ & $ \p_j$ & $\p_{jj}$ & $\frac{\p_{jl} + \p_{jl}}{\sqrt{2}}$ \\
		\hline
		$\p_i $ & $a_1 \, \delta_{ij}$ & $0$ & $0$\\ 
		$\p_{ii}$ & $0$ & $\left[ (a_2 + a_3)\, \delta_{ij} \delta_{ij} + a_4 \, \delta_{ii} \delta_{jj} \right] l_s^{-2} $ & $0$  \\
		$\frac{\p_{ik}+\p_{ki}}{\sqrt{2}}$ & $0$ & $0$ & $(a_2 + a_3) \, \delta_{ij} \delta_{kl}\, l_s^{-2}$  
	\end{tabular}
\end{equation}
We note that the metric splits the tangent space in three orthogonal subspaces: one spanned by $\{\p_i \, | \, i\in\{ 1,...,n\}  \}$, a second spanned by $\{\p_{ii} \, |\, i\in\{ 1,...,n\}  \}$ and a third by $\{(\p_{ij}+\p_{ji})/\sqrt{2} \, |\, i<j\in\{ 1,...,n\}  \}$. The first and third are themselves orthogonal, and can be made orthonormal by fixing
\begin{align}
	a_1 &= 1 \, ,\nonumber\\
	a_2 + a_3 &= 1 \, .
\end{align}
We will implement the second equation by setting
\begin{equation}
	a_2 = \frac{1 + c_1}{2}\, , \qquad a_3 = \frac{1 - c_1}{2} \, ,  \qquad c_1\in\R \, ,
\end{equation}
where $c_1$ is a free parameter, which is irrelevant for our purpose, as the symmetrized metric $G_{(^\mu_{\rho\sigma})(^\nu_{\kappa\lambda})}$ does not depend on it.
The second space is orthogonal only if $a_3=0$, such that $a_3$ measures the deviation from orthogonality. We will leave this as a free parameter, but redefine the parameter as
\begin{equation}
	a_3 = \frac{c_2-1}{n} \, \qquad c_2\in\R.
\end{equation}
We conclude that the flat space metric is given by
\begin{equation}
	\begin{pmatrix}
		G_{ij} & G_{i|mn|} \\
		G_{|kl|j} & G_{|kl|mn|}
	\end{pmatrix}
	= 
	\begin{pmatrix}
		\delta_{ij} & 0 \\
		0 &  l_s^{-2} \left[ \frac{1+c_1}{2} \, \delta_{km} \delta_{ln} + \frac{1-c_1}{2} \, \delta_{kn} \delta_{lm} 
		- \frac{1-c_2}{n} \, \delta_{kl} \delta_{mn} \right]
	\end{pmatrix} 
\end{equation}
and the cometric by
\begin{equation}
	\begin{pmatrix}
		G^{ij} & G^{i|mn|} \\
		G^{|kl|j} & G^{|kl|mn|}
	\end{pmatrix}
	= 
	\begin{pmatrix}
		\delta^{ij} & 0 \\
		0 & l_s^2 \left[ \frac{1+ c_1}{2 \, c_1} \delta^{km} \delta^{ln} - \frac{1-c_1}{2 \, c_1} \delta^{kn} \delta^{lm}  + \frac{(1-c_2)}{n \,c_2} \, \delta^{kl} \delta^{mn} \right]
	\end{pmatrix}.
\end{equation}
By calculating the eigenvalues of the metric, one finds the signature of the second order tangent space. The eigenvalues are given by
\begin{center}
	\begin{tabular}{ c | c }
		Eigenvalue & Multiplicity \\
		\hline
		$1$ & $N-1$\\
		$c_2$ & $1$
	\end{tabular}
\end{center}
where $N$ is the dimension of the second order tangent space, as given in eq.~\eqref{eq:dimTangentsecond}.
Thus, for $c_2>0$ the second order metric has a Euclidean signature, for $c_2<0$ the metric has a Lorentzian signature, and for $c_2=0$ the metric is degenerate.
\par 

We will continue with determining the metric on the second order tangent bundle for the Minkowski spacetime $\R^{n-1,1}$. This can done following the same analysis as before, and the result can be obtained by replacing the Kronecker delta with the Minkowski metric. Hence, the metric is given by
\begin{equation}
	\begin{pmatrix}
		G_{\mu\nu} & G_{\mu|\kappa\lambda|} \\
		G_{|\rho\sigma|\nu} & G_{|\rho\sigma|\kappa\lambda|}
	\end{pmatrix}
	= 
	\begin{pmatrix}
		\eta_{\mu\nu} & 0 \\
		0 & l_s^{-2} \left[ \frac{1+c_1}{2} \, \eta_{\rho\kappa} \eta_{\sigma\lambda} + \frac{1-c_1}{2} \eta_{\rho\lambda} \eta_{\sigma\kappa}  - \frac{1-c_2}{n} \, \eta_{\rho\sigma} \eta_{\kappa\lambda} \right]
	\end{pmatrix}
\end{equation}
and the cometric by
\begin{equation}
	\begin{pmatrix}
		G^{\mu\nu} & G^{\mu|\kappa\lambda|} \\
		G^{|\rho\sigma|\nu} & G^{|\rho\sigma|\kappa\lambda|}
	\end{pmatrix}
	= 
	\begin{pmatrix}
		\eta^{\mu\nu} & 0 \\
		0 & l_s^2 \left[ \frac{1+c_1}{2\, c_1} \, \eta^{\rho\kappa} \eta^{\sigma\lambda} - \frac{1-c_1}{2\, c_1} \,  \eta^{\rho\lambda} \eta^{\sigma\kappa}  + \frac{1-c_2}{n \, c_2}\, \eta^{\rho\sigma} \eta^{\kappa\lambda} \right]
	\end{pmatrix}.
\end{equation}
In order to determine the signature, we calculate the eigenvalues and find
\begin{center}
	\begin{tabular}{ c | c }
		Eigenvalue & Multiplicity \\
		\hline
		$1$ & $N-1-n$\\
		$-1$ & $n$ \\
		$c_2$ & $1$
	\end{tabular}
\end{center}

\subsection{Curved space(time)}
In this subsection, we extend the metric on flat space(time) to curved spacetime. A part of this metric can by obtained by replacing the Euclidean metric $\delta$ or Minkowski metric $\eta$ with a generic (pseudo-)Riemannian metric $g$. However, further corrections associated to the derivatives of the (co)metric may be obtained. In order to estimate these corrections, we will first evaluate the inner product of the covariant basis elements, as constructed in eqs.~\eqref{eq:CovVec} and \eqref{eq:CovForm}. This yields
\begin{center}
	\begin{tabular}{ c | c c }
		$\cdot$ & $ \hat{\p}_\nu$ & $\hat{\p}_{\kappa\lambda}$ \\
		\hline
		$\hat{\p}_\mu$ 
		& $\hat{G}_{\mu\nu}$ 
		& $\hat{G}_{\mu|\kappa\lambda|}$ 
		\\ 
		$\hat{\p}_{\rho\sigma} $ 
		& $\hat{G}_{|\rho\sigma|\nu}$ 
		& $\hat{G}_{|\rho\sigma|\kappa\lambda|}$   
	\end{tabular}
\end{center}
where $\hat{G}$ can be expressed in terms of $G$ as follows
\begin{align}
	\hat{G}_{\mu\nu} &= G_{\mu\nu} \, ,\\
	\hat{G}_{\mu|\kappa\lambda|} &= G_{\mu|\kappa\lambda|}  - \frac{1}{2} \, G_{\mu\nu} \Gamma^\nu_{\kappa\lambda} \, \\
	\hat{G}_{|\rho\sigma|\nu} &= G_{|\rho\sigma|\nu}  - \frac{1}{2} \, \Gamma^\mu_{\rho\sigma} G_{\mu\nu} \, ,\\
	\hat{G}_{|\rho\sigma|\kappa\lambda|} &= G_{|\rho\sigma|\kappa\lambda|} - \frac{1}{2} \, G_{|\rho\sigma|\nu} \Gamma^\nu_{\kappa\lambda} 
	- \frac{1}{2} \, \Gamma^\mu_{\rho\sigma} G_{\mu|\kappa\lambda|}
	+ \frac{1}{4} \, \Gamma^\mu_{\rho\sigma} G_{\mu\nu} \Gamma^\nu_{\kappa\lambda}
\end{align}

Similarly, on the covariant cotangent basis, one obtains
\begin{center}
	\begin{tabular}{ c | c c }
		$\cdot$ & $ \rd_2 \hat{x}^\nu$ & $\rd\langle \hat{x}^\kappa,\hat{x}^\lambda \rangle$ \\
		\hline
		$\rd_2\hat{x}^\mu$ 
		& $\hat{G}^{\mu\nu}$ 
		& $\hat{G}^{\mu|\kappa\lambda|}$ 
		\\ 
		$\rd\langle \hat{x}^\rho ,\hat{x}^\sigma\rangle$ 
		& $\hat{G}^{|\rho\sigma|\nu}$ 
		& $\hat{G}^{|\rho\sigma|\kappa\lambda|} $   
	\end{tabular}
\end{center}
with
\begin{align}
	\hat{G}^{\mu\nu} 
	&= G^{\mu\nu} 
	+ \frac{1}{2} \, G^{\mu|\kappa\lambda|} \Gamma^\nu_{\kappa\lambda} 
	+ \frac{1}{2} \, \Gamma^\mu_{\rho\sigma} G^{|\rho\sigma|\kappa\lambda|}
	+ \frac{1}{4} \, \Gamma^\mu_{\rho\sigma} G^{|\rho\sigma|\kappa\lambda|} \Gamma^\nu_{\kappa\lambda} 
	\, ,\\
	\hat{G}^{\mu|\kappa\lambda|}
	&= 
	G^{\mu|\kappa\lambda|} 
	+ \frac{1}{2} \, \Gamma^\mu_{\rho\sigma} G^{|\rho\sigma|\kappa\lambda|} 
	\, ,\\
	\hat{G}^{|\rho\sigma|\nu} 
	&=
	G^{|\rho\sigma|\nu} 
	+ \frac{1}{2} \, G^{|\rho\sigma|\kappa\lambda|} \Gamma^\nu_{\kappa\lambda} 
	\, \\
	\hat{G}^{|\rho\sigma|\kappa\lambda|}
	&=
	G^{|\rho\sigma|\kappa\lambda|} \, .
\end{align}
\par 

We can now evaluate the inner product in terms of the covariant metric elements
\begin{align}
	G(u,v) 
	&= G_{\mu\nu} u^\mu v^\nu 
	+ G_{\mu|\kappa\lambda|} u^\mu v_2^{\kappa\lambda} 
	+ G_{|\rho\sigma|\nu} u_2^{\rho\sigma} v^\nu 
	+ G_{|\rho\sigma|\kappa\lambda|} u_2^{\rho\sigma} v_2^{\kappa\lambda} \nonumber\\
	&= \hat{G}_{\mu\nu} \hat{u}^\mu \hat{v}^\nu 
	+ \hat{G}_{\mu|\kappa\lambda|}\hat{u}^\mu \hat{v}_2^{\kappa\lambda} 
	+ \hat{G}_{|\rho\sigma|\nu} \hat{u}_2^{\rho\sigma} \hat{v}^\nu 
	+ \hat{G}_{|\rho\sigma|\kappa\lambda|} \hat{u}_2^{\rho\sigma} \hat{v}_2^{\kappa\lambda}  \, .
\end{align}
Then, using that the inner product must be invariant under coordinate transformations, that $\hat{v}^\mu$ transforms as a first order vector, and that $\hat{v}_2^{\mu\nu}$ transforms as a first order (2,0)-tensor, one finds that $\hat{G}_{\mu\nu}$ should transform as a first order (0,2)-tensor, $G_{\mu|\kappa\lambda|}$ and $G_{|\rho\sigma|\nu}$ should transform as first order $(0,3)$-tensors, and $\hat{G}_{|\rho\sigma|\kappa\lambda|}$ should transform as a first order $(0,4)$-tensor. Therefore, any correction to the metric $\hat{G}$ should involve tensorial structures constructed from the derivatives of the first order (co)metric. 
\par 

The first order sector does not contain any derivatives of the (co)metric, such that it remains unchanged. Hence,
\begin{equation}
	\hat{G}_{\mu\nu} = g_{\mu\nu} \, .
\end{equation}
The off-diagonal terms can be corrected by tensorial structures that are constructed using first order derivatives of the (co)metric. There are two such structures: the torsion tensor $T$ and the non-metricity tensor $Q$. Hence, if we assume the first order metric to be metric compatible and torsion-free, one obtains
\begin{align}
	\hat{G}_{\mu|\kappa\lambda|} &= 0 \, ,\\
	\hat{G}_{|\rho\sigma|\nu} &= 0 \, .
\end{align}
Finally, the second order sector can be modified by tensorial structures obtained from second order derivatives of the first order (co)metric. Hence,
\begin{align}\label{eq:CovMetric}
	\hat{G}_{|\rho\sigma|\kappa\lambda|} &= 
	l_s^{-2} \left[ \frac{1+c_1}{2} \, g_{\rho\kappa} g_{\sigma\lambda} + \frac{1-c_1}{2} \, g_{\rho\lambda} g_{\sigma\kappa} - \frac{1-c_2}{n} \, g_{\rho\sigma} g_{\kappa\lambda} \right] 
	+ A_{|\rho\sigma|\kappa\lambda|} \, ,
\end{align}
where $A$ is a first order $(0,4)$-tensor. If we assume the first order metric to be metric compatible and torsion-free, it can only be corrected by the curvature tensor $\mathcal{R}$, such that it is given by
\begin{align}\label{eq:MetricATen}
	A_{|\rho\sigma|\kappa\lambda|} 
	&= 
	b_1 \, \mathcal{R}_{\rho\kappa\sigma\lambda} 
	+ b_2 \,  \mathcal{R}_{\rho\lambda\sigma\kappa}
	+ \left( b_3 \, g_{\rho\kappa} g_{\sigma\lambda} + b_4 \, g_{\rho\lambda} g_{\sigma\kappa} + b_5 \, g_{\rho\sigma} g_{\kappa\lambda} \right) \mathcal{R}
	\nonumber\\
	&\quad
	+ b_6 \left( g_{\rho\sigma} \mathcal{R}_{\kappa\lambda} + g_{\kappa\lambda} \mathcal{R}_{\rho\sigma} \right) 
	+ b_7 \left( g_{\rho\lambda} \mathcal{R}_{\sigma\kappa} + g_{\sigma\kappa} \mathcal{R}_{\rho\lambda} \right) 
	+ b_8 \, g_{\rho\kappa} \mathcal{R}_{\sigma\lambda}
	+ b_9 \, g_{\sigma\lambda} \mathcal{R}_{\rho\kappa},
\end{align}
where $b_i\in\R$ must be determined by further calculations. As the second order parts of vectors are symmetric, only the symmetrized metric is relevant fr our purposes. After symmetrization, one obtains
\begin{align}
	\hat{G}_{(\rho\sigma)(\kappa\lambda)} &= 
	l_s^{-2} \left[ \frac{1}{2} \, g_{\rho\kappa} g_{\sigma\lambda} + \frac{1}{2} \, g_{\rho\lambda} g_{\sigma\kappa} - \frac{1-c_2}{n} \, g_{\rho\sigma} g_{\kappa\lambda} \right] 
	+ A_{(\rho\sigma)(\kappa\lambda)} \, ,
\end{align}
with 
\begin{align}
	A_{(\rho\sigma)(\kappa\lambda)} 
	&= 
	\frac{b_1 +b_2}{2} \left( \mathcal{R}_{\rho\kappa\sigma\lambda} 
	+ \mathcal{R}_{\rho\lambda\sigma\kappa} \right)
	+ \left[ \frac{b_3+ b_4}{2} \left( g_{\rho\kappa} g_{\sigma\lambda} + g_{\rho\lambda} g_{\sigma\kappa} \right) + b_5 \, g_{\rho\sigma} g_{\kappa\lambda} \right] \mathcal{R}
	\\
	&\quad
	+ b_6 \left( g_{\rho\sigma} \mathcal{R}_{\kappa\lambda} + g_{\kappa\lambda} \mathcal{R}_{\rho\sigma} \right) 
	+ \frac{2\,b_7 + b_8 + b_9}{4}  \left( g_{\rho\lambda} \mathcal{R}_{\sigma\kappa} + g_{\sigma\kappa} \mathcal{R}_{\rho\lambda} + g_{\rho\kappa} \mathcal{R}_{\sigma\lambda}
	+ g_{\sigma\lambda} \mathcal{R}_{\rho\kappa} \right),\nonumber
\end{align}

We note that the correction $A_{|\rho\sigma|\kappa\lambda|}$ generates the Pauli-DeWitt term \cite{DeWitt:1957,Pauli:1973}, which arises as a correction to the Hamiltonian for quantum theories on curved space. After contraction with the metric, one obtains
\begin{equation}
	g^{\rho\sigma} A_{|\rho\sigma|\kappa\lambda|} g^{\kappa\lambda} 
	= \left[ \left( b_1 + b_2 \right) + n \, (b_3 + b_4 + n \, b_5) + \left( 2 \, n \, b_6 + 2\, b_7 + b_8 + b_9\right) \right] \mathcal{R}
	=: c_3 \, \mathcal{R}
\end{equation}
The value calculated by DeWitt in Ref.~\cite{DeWitt:1957} can be reproduced by setting $c_3=\frac{1}{6}$. This particular value has been confirmed by several other calculations, for example within stochastic mechanics \cite{Nelson:1985,Kuipers:2021ylr,Kuipers:2023pzm}. However, DeWitt corrected his calculation in later work \cite{DeWitt:2003pm,DeWitt:2012mdz}, and this corrected result can be reproduced by setting $c_3=\frac{1}{4}$. As the value of $c_3$ is debated, we will leave it as a free $O(1)$-parameter and leave its calculation in the framework of second order geometry for future work.
\par 

From eq.~\eqref{eq:CovMetric} it is clear that the correction due to the Riemann term will not alter the signature of the metric as long as the curvature scalars, such as the Ricci and Kretschmann scalar are bounded by this length scale, i.e.  $\mathcal{R},|\mathcal{R}_{\mu\nu\rho\sigma} \mathcal{R}^{\mu\nu\rho\sigma}|^{1/2} \ll l_s^{-2}$. Therefore, the length scale $l_s$ should be taken small, and the resulting theory is well-behaved as long as the curvature remains small compared to the inverse of this length scale. Beyond that scale second order geometry is still valid, but there no longer exists a well-behaved second order metric.\footnote{Note that the choice of $c_2$ may also alter the metric signature. However, this is a global choice, thus fixed on the entire manifold. Large curvature, on the other hand, may change the signature of the second order metric locally.}
\par 

The cometric can be obtained by inverting the metric. This yields
\begin{align}
	\hat{G}^{\mu\nu} &= g^{\mu\nu} \, \nonumber\\
	\hat{G}^{\mu|\kappa\lambda|} &= 0 \, ,\nonumber\\
	\hat{G}^{|\rho\sigma|\nu} &= 0 \, ,\nonumber\\
	\hat{G}^{|\rho\sigma|\kappa\lambda|} &= 
	l_s^2 \, \sum_{k=0}^\infty (- l_s^2)^{k} \, \hat{G}_{k}^{|\rho\sigma|\kappa\lambda|} \, ,
\end{align}
with
\begin{align}
	\hat{G}_0^{|\rho\sigma|\kappa\lambda|}
	&= 	\frac{1 + c_1}{2 \, c_1} \, g^{\rho\kappa} g^{\sigma\lambda} 
		- \frac{1 - c_1}{2 \, c_1} \, g^{\rho\lambda} g^{\sigma\kappa} 
		+ \frac{1 - c_2}{n \, c_2} \, g^{\rho\sigma} g^{\kappa\lambda}\, , \\
	\hat{G}_{k}^{|\rho\sigma|\kappa\lambda|}
	&= \hat{G}_0^{|\rho\sigma|\alpha\beta|} 
	A_{|\alpha\beta|\gamma\delta|}
	\hat{G}_{k-1}^{|\gamma\delta|\kappa\lambda|} \qquad k\in\mathbb{N}\, .
\end{align}
\par 

We conclude this section by noting that the additional $\hat{G}_{|\rho\sigma|\kappa\lambda|}$ can be interpreted as an area metric, as was already mentioned in section \ref{sec:Intro}. An area metric is a function $\mathfrak{G}:(T_x\M)^{\otimes4}\rightarrow \R$ with the following symmetries \cite{Schuller:2005yt}
\begin{align}
	\mathfrak{G}(X,Y,U,V) &= \mathfrak{G}(U,V,X,Y) \, ,\\
	\mathfrak{G}(X,Y,U,V) &= - \mathfrak{G}(X,Y,V,U) \, ,\label{eq:MetricAntiSymm}\\
	\mathfrak{G}(X,Y,U,V) + \mathfrak{G}(X,U,V,Y) + \mathfrak{G}(X,V,Y,U) &= 0 \, .
\end{align}
We may now impose these symmetries on $\hat{G}_{|\rho\sigma|\kappa\lambda|}$. However, this would require an extension of the map $\mathcal{H}$, given in eq.~\eqref{eq:MapH} to bilinear second order forms. There are various possibilities associated to the permutations of  $(\rho\sigma\kappa\lambda)$. A first option would be 
\begin{equation}
	\hat{G}_{|\rho\sigma|\kappa\lambda|} \sim \mathfrak{G}_{\rho\sigma\kappa\lambda}  \, ,
\end{equation}
but in that case the anti-symmetry relation \eqref{eq:MetricAntiSymm} is inconsistent with our expression of $G_{|\rho\sigma|\kappa\lambda|}$. This shows that such an extension of $\mathcal{H}$ is incompatible with an area metric interpretation of $G$. Alternatively, the extension of $\mathcal{H}$ could be such that
\begin{equation}
	G_{|\rho\sigma|\kappa\lambda|} \sim \mathfrak{G}_{\rho\kappa\sigma\lambda} \, .
\end{equation}
In this case, the symmetries of the area metric impose
\begin{align}
	\hat{G}_{|\rho\sigma|\kappa\lambda|} &= \hat{G}_{|\sigma\rho|\lambda\kappa|} \, ,\\
	\hat{G}_{|\rho\sigma|\kappa\lambda|} &= - \hat{G}_{|\rho\lambda|\kappa\sigma|} \, ,\\
	\hat{G}_{|\rho\sigma|\kappa\lambda|} + \hat{G}_{|\rho\kappa|\lambda\sigma|} + \hat{G}_{|\rho\lambda|\sigma\kappa|} &= 0 \, .
\end{align} 
These symmetries impose various constraints on the free parameters 
\begin{alignat}{2}
	c_1 &= -1 \, , \nonumber\\
	c_2 &= \frac{2 - n + n \, c_1}{2} &&= 1 - n \, ,
\end{alignat}
and
\begin{alignat}{2}
	b_2 &= 0 \, ,\nonumber\\
	b_3 &= 0 \, ,\nonumber\\
	b_5 &= - b_4 \, , \nonumber\\
	b_7 &= - b_6 \, , \nonumber\\
	b_9 &= b_8 &&=0 \, ,
\end{alignat}
such that the metric is given by
\begin{align}
	\hat{G}_{|\rho\sigma|\kappa\lambda|} 
	&= 
	\frac{g_{\rho\lambda} g_{\sigma\kappa} - g_{\rho\sigma} g_{\kappa\lambda}}{l_s^{2}} 
	+ b_1 \, \mathcal{R}_{\rho\kappa\sigma\lambda}
	+ b_4 \left( g_{\rho\lambda} g_{\sigma\kappa} - g_{\rho\sigma} g_{\kappa\lambda} \right) \mathcal{R}
	\nonumber\\
	&\quad
	+ b_6 \left( g_{\rho\sigma} \mathcal{R}_{\kappa\lambda} + g_{\kappa\lambda} \mathcal{R}_{\rho\sigma} - g_{\rho\lambda} \mathcal{R}_{\sigma\kappa} - g_{\sigma\kappa} \mathcal{R}_{\rho\lambda} \right)  
\end{align}
and
\begin{equation}
	c_3
	= b_1 - n \, (n - 1 ) \, b_4  + 2 \, (n - 1 ) \,  b_6 \, .
\end{equation}
Hence, by imposing an area metric interpretation, one can further constrain the second order metric.

\section{Construction of physical theories}\label{eq:PhysTheory}
Physical theories can be conveniently described in a Lagrangian formalism. In the classical  theory, this can be used to derive the equations of motion of the system directly. In particular, in the case of a single scalar particle, one can obtain its trajectory by extremizing an action
\begin{equation}\label{eq:FirstOrderAction}
	S(X,\mathcal{T}) = \int_\mathcal{T} L[X(\tau),\dot{X}(\tau),\tau]  \, d\tau \, ,
\end{equation}
where $L:T\M \times \mathcal{T} \rightarrow \R$ is a function on the tangent bundle. In the quantum theory, on the other hand, the action appears in the probability measure of the path integral. Indeed, the probability that a particle is localized on a specific trajectory $X(\omega,\cdot):\mathcal{T}\rightarrow\M$, labeled by $\omega\in\Omega$, is described by eq.~\eqref{eq:PathIntMeasure}:
\begin{equation*}
	\rho(X,\mathcal{T}) \sim e^{\ri S(X,\mathcal{T})} = e^{\ri \int_\mathcal{T} L(X,\dot{X},\tau) \, d\tau} \, .
\end{equation*}
However, as pointed out in section \ref{sec:PIProp}, the differentiable paths have zero measure in the path integral, such that $\dot{X}$ is not well-defined along the paths in the path integral. Nevertheless, as pointed out in section \ref{sec:Differentials}, there exist well-defined velocity fields $v_\pm(X(\tau),\tau)$, such that the action is of the form
\begin{equation}
	S_\pm(X,\mathcal{T}) = \int_\mathcal{T} L[X(\tau),v(X(\tau),\tau),\tau]  \, d\tau \, .
\end{equation}
However, these velocity fields $v_\pm$ do not encode all relevant information, as there are also second order velocity fields $v_{2\pm}$. Therefore, the Lagrangian should be defined on the second order tangent bundle, i.e. $L:T_2\M \times \mathcal{T} \rightarrow \R$ with an action given by
\begin{equation}
	S_\pm(X,\mathcal{T}) = \int_\mathcal{T} L[X(\tau),v(X(\tau),\tau), v_{2}(X(\tau),\tau),\tau]  \, d\tau \, .
\end{equation}
\par 

On a flat spacetime the first and second order parts of vectors are disentangled, such that the Lagrangian can be decomposed as
\begin{equation}
	L_{\rm flat}(x,v,v_{2},\tau)  
	= L_{1}(X,v,\tau) + L_{2}(X, v_{2},\tau) \, ,
\end{equation}
and the probability measure \eqref{eq:PathIntMeasure} can be split as
\begin{equation}\label{eq:LagSplit}
	\rho_{\rm flat}(X,\mathcal{T}) \sim e^{ \ri \int_\mathcal{T} L_1(X,v,\tau) \, d\tau} \, e^{ \ri \int_\mathcal{T} L_2(X, v_{2},\tau) \, d\tau} \; .
\end{equation}
Thus, on a flat spacetime the second order part affects the normalization of the path integral, but the dynamics of the theory can still be described using the first order Lagrangian \eqref{eq:FirstOrderAction}. 
On a curved spacetime, on the other hand, the first and second order parts of vectors are coupled under general coordinate transformations, as shown in eq.~\eqref{eq:Vec2Trans}. Therefore, one cannot split the Lagrangian as in eq.~\eqref{eq:LagSplit}. Therefore, a path integral formulation based on the first order action \eqref{eq:FirstOrderAction} is expected to generate anomalies for quantum theories on curved spacetime.

\subsection{Second order Lagrangian and Hamiltonian}

Given the importance of the second order Lagrangian, we are faced with the problem of constructing a second order Lagrangian from a first order Lagrangian. In order to do this, we will consider a generic first order Lagrangian for both a non-relativistic and relativistic theory. These are given by
\begin{align}
	L_{\rm NRT}(x,v,t) &= \frac{m}{2} \, g_{ij}(x,t) \, v^i v^j + q \, A_i(x,t) \, v^i - \mathfrak{U}(x,t) \, ,\\
	L_{\rm RLT}(x,v) &= \frac{1}{2\, \varepsilon(\tau)} \, g_{\mu\nu}(x) \, v^\mu v^\nu + q \, A_\mu(x) \, v^\mu  - \frac{\varepsilon(\tau) \, m^2}{2} \, ,
\end{align}
where $\varepsilon$ is gauge fixed in the equations of motion by\footnote{This condition should be imposed in the equations of motion, but not in the Lagrangian, cf. e.g. \cite{Lust:1989tj}.}
\begin{equation}\label{eq:Epsilonconditon}
	\varepsilon(\tau) = 
	\begin{cases}
		m^{-1} \qquad &\textrm{if} \quad m>0 \,;\\
		E^{-1}(\tau) \qquad &\textrm{if} \quad m=0 \,.
	\end{cases}
\end{equation}
The second order Lagrangian corresponding to these Lagrangians can be obtained using the results from previous sections: the first term will be generalized using the second order metric discussed in section \ref{sec:Metric}; the second term can be generalized using the map $\underline{\rd}$ discussed in eq.~\eqref{eq:d2f}; the third term is unchanged, as it does not depend on the velocity. Therefore, the second order Lagrangians are given by
\begin{align} \label{eq:Lag2NRT}
	L_{\rm NRT}(x,v,v_2,t) 
	&= \frac{m}{2} \left[  g_{ij} \hat{v}^i \hat{v}^j
	+ \hat{G}_{|ij|kl|} \hat{v}_2^{ij} \hat{v}_2^{kl}
	\right]
	+ q \, A_i \hat{v}^i
	+ \frac{q}{2} \, \nabla_j A_i \hat{v}_2^{ij} 
	- \mathfrak{U} \, ,\\
	L_{\rm RLT}(x,v,v_2) 
	&= \frac{1}{2\, \varepsilon} \left[ 
	g_{\mu\nu} \hat{v}^\mu \hat{v}^\nu 
		+ \hat{G}_{|\mu\nu|\rho\sigma|} \hat{v}_2^{\mu\nu} \hat{v}_2^{\rho\sigma}
	\right]
	+ q \, A_\mu \hat{v}^\mu + \frac{q}{2} \, \nabla_\nu A_\mu \hat{v}_2^{\mu\nu} 
	- \frac{\varepsilon \, m^2}{2}  \,, \label{eq:Lag2RLT}
\end{align}
where the second order metric is given in eq.~\eqref{eq:CovMetric}.
\par 

Using these Lagrangians, one can define momenta by 
\begin{align}
	p_\mu &:= \frac{\p L}{\p v^\mu} \, , \\
	p^{(2)}_{\mu\nu} &:= \frac{\p L}{\p v_{2}^{\mu\nu}} \, ,
\end{align}
One may also define covariant momenta by
\begin{alignat}{4}
	\hat{p}_\mu 
	&:= \frac{\p L}{\p \hat{v}^\mu} 
	&&= \frac{\p v^\nu}{\p \hat{v}^\mu} \frac{\p L}{\p v^\nu}
	+ \frac{\p v_2^{\nu\rho}}{\p \hat{v}^\mu} \frac{\p L}{\p v_2^{\nu\rho}}
	&&= p_\mu \, ,\\
	\hat{p}^{(2)}_{\mu\nu} &:= \frac{\p L}{\p \hat{v}_{2}^{\mu\nu}} 
	&&= \frac{\p v^\rho}{\p \hat{v}_{2}^{\mu\nu}} \frac{\p L}{\p v^\rho}
	+ \frac{\p v_2^{\rho\sigma}}{\p \hat{v}_{2}^{\mu\nu}} \frac{\p L}{\p v_{2}^{\rho\sigma}} 
	&&= p^{(2)}_{\mu\nu} - \frac{1}{2} \, \Gamma^\rho_{\mu\nu} p_\rho	\, ,
\end{alignat}
which is consistent with the definition of covariant vectors \eqref{eq:CovVec} and forms \eqref{eq:CovForm}.
Finally, using the momenta, one can construct a second order Hamiltonian by \cite{Huang:2022}
\begin{align}
	H(x,p,p_2) 
	&=  p_\mu v^\mu  
	+ p^{(2)}_{\mu\nu} v_{2}^{\mu\nu} 
	- L(x,v,v_{2}) 
	\nonumber\\
	&=  \hat{p}_\mu \hat{v}^\mu  
	+ \hat{p}^{(2)}_{\mu\nu} \hat{v}_{2}^{\mu\nu} 
	- L(x,v,v_{2})\, .
\end{align}
Then, for the given Lagrangians \eqref{eq:Lag2NRT} and \eqref{eq:Lag2RLT} one obtains the covariant relativistic momenta
\begin{align}
	\hat{p}_\mu &= \varepsilon^{-1} \, g_{\mu\nu}\hat{v}^\nu + q \, A_\mu \, ,\\
	\hat{p}^{(2)}_{\mu\nu} &=  \varepsilon^{-1} \, \hat{G}_{|\mu\nu|\rho\sigma|} \hat{v}_{2}^{\rho\sigma} + \frac{q}{2} \, \nabla_{\nu} A_\mu \, ,
\end{align}
which can easily be generalized to the non-relativistic case.
Finally, the Hamiltonian is given by 
\begin{align}\label{eq:Hamiltonian2NRT}
	H_{\rm NRT}(x,p,p_2,t) 
	&= 
	\frac{1}{2\, m} \left[ g^{ij} \left( \hat{p}_i - q\, A_i \right) \left( \hat{p}_j - q\, A_j \right)
	+ \hat{G}^{|ij|kl|} \left(\hat{p}^{(2)}_{ij} - \frac{q}{2} \, \nabla_j A_i \right) \left(\hat{p}^{(2)}_{kl} - \frac{q}{2}\, \nabla_l A_k \right)
	\right]
	+ \mathfrak{U} \, ,\\
	H_{\rm RLT}(x,p,p_2) 
	&= 
	\frac{\varepsilon}{2} \left[ g^{\mu\nu} \left( \hat{p}_\mu - q\, A_\mu \right) \left( \hat{p}_\nu - q\, A_\nu \right)
	+ \hat{G}^{|\mu\nu|\rho\sigma|} \left(\hat{p}^{(2)}_{\mu\nu} - \frac{q}{2} \, \nabla_\nu A_\mu \right) \left(\hat{p}^{(2)}_{\rho\sigma} - \frac{q}{2}\, \nabla_\sigma A_\rho \right)
	+ m^2
	\right] .\label{eq:Hamiltonian2RLT}
\end{align}

\subsection{Two velocity fields}\label{sec:2Vels}
A second problem, that arises in constructing an action, that can be used in the path integral formulation, is the presence of two linearly independent velocity fields $(v,v_{2})_+$ and $(v,v_{2})_-$ or equivalently  $(v,v_{2})_\circ$ and $(v,v_{2})_\perp$. This leads to an ambiguity in the definition of the action, as it is not clear which one should be chosen, cf. e.g. Ref.~\cite{DeWitt:1957}. In order to avoid the ambiguity, we may construct a more general theory that depends on both velocity fields. There are various options for doing so \cite{Nelson:1985}: one option would be to take a linear combination of a Lagrangian depending on $v_+$ and a Lagrangian depending on $v_-$ \cite{Yasue:1981}; another option would be to consider a single Lagrangian depending on a velocity field $\mathfrak{v}$ that is a linear combination of $v_+$ and $v_-$ \cite{Guerra:1982fn}. Here, we will generalize these two options by considering the linear combination of two Lagrangians that depend on velocity fields $\mathfrak{v}_{1,2}$ that are both linear combinations of $v_+$ and $v_-$. Furthermore, we will subject the resulting Lagrangian to the constraint that in the classical limit, where $v_{+}=v_-$, the first order Lagrangian is obtained. 
\par 

We will thus consider the following action
\begin{equation}
	S(X,\mathcal{T}) =  \int_\mathcal{T} L(x,v_\circ,v_\perp,v_{2\circ},v_{2\perp}) \, d\tau  
\end{equation}
with Lagrangian
\begin{equation}\label{eq:Lagrangian0}
	L(x,v_\circ,v_\perp,v_{2\circ},v_{2\perp}) =
	\frac{1 + \alpha_3}{2} \, L(x,\mathfrak{v}_1)
	+ \frac{1 - \alpha_3}{2} \, L(x,\mathfrak{v}_2) \, ,
\end{equation}
where
\begin{equation}
	\mathfrak{v}_{1,2} = 
	\begin{pmatrix}
		v_\circ \\
		v_{2\circ}
	\end{pmatrix}
	+ \alpha_{1,2}
	\begin{pmatrix}
		v_\perp \\
		v_{2\perp}
	\end{pmatrix}
\end{equation}
and $\alpha_1,\alpha_2,\alpha_3\in\mathbb{C}$. Note that we allow for complex parameters as would be expected in a quantum theory \cite{Pavon:1995April,Pavon:2000,Kuipers:2023pzm,Kuipers:2023ibv}.
\par 

We will focus on the relativistic theory\footnote{The non-relativistic theory can easily be obtained by replacing $\mu,\nu,...\rightarrow i,j,..$, $\varepsilon\rightarrow m^{-1}$ and $\frac{\varepsilon m^2}{2}\rightarrow \mathfrak{U}$.} and we will work in the $(v_\circ,v_\perp)$ basis\footnote{The results in the $(v_+,v_-)$ basis or any other basis can easily be obtained from this using that $v_\circ=(v_++v_-)/2$ and $v_\perp=(v_+-v_-)/2$.}. The Lagrangian is then given by
\begin{align}\label{eq:LagRLT2Vels}
	L_{\rm RLT} 
	&=
	\frac{1}{2\, \varepsilon} \left[ 
	g_{\mu\nu} \hat{v}_\circ^\mu \hat{v}_\circ^\nu 
	+ \hat{G}_{|\mu\nu|\rho\sigma|} \hat{v}_{2\circ}^{\mu\nu} \hat{v}_{2\circ}^{\rho\sigma}
	\right]
	+ \frac{\beta_2^2}{2\, \varepsilon} \left[ 
	g_{\mu\nu} \hat{v}_\perp^\mu \hat{v}_\perp^\nu 
	+ \hat{G}_{|\mu\nu|\rho\sigma|} \hat{v}_{2\perp}^{\mu\nu} \hat{v}_{2\perp}^{\rho\sigma}
	\right]
	\nonumber\\
	&\quad
	+ \frac{\beta_1}{\varepsilon} \left[ 
	g_{\mu\nu} \hat{v}_\circ^\mu \hat{v}_\perp^\nu 
	+ \hat{G}_{|\mu\nu|\rho\sigma|} \hat{v}_{2\circ}^{\mu\nu} \hat{v}_{2\perp}^{\rho\sigma}
	\right]
	+ q \left( A_\mu v_\circ^\mu 
	+ \frac{1}{2} \, \nabla_\nu A_\mu v_{2\circ}^{\mu\nu} \right)
	\nonumber\\
	&\quad
	+ \beta_1 \, q
	\left(  A_\mu v_\perp^\mu + \frac{1}{2} \, \nabla_\nu A_\mu v_{2\perp}^{\mu\nu} \right)
	- \frac{\varepsilon \, m^2}{2} \, ,
\end{align}
where
\begin{align}\label{eq:betadef}
	\beta_1 &= \frac{\left(\alpha_1 + \alpha_2\right)+\left(\alpha_1 - \alpha_2\right) \alpha_3}{2} \, ,\nonumber\\
	\beta_2^2 &= \frac{\left(\alpha_1^2 + \alpha_2^2\right)+\left(\alpha_1^2 - \alpha_2^2\right) \alpha_3}{2} \, .
\end{align}

\subsubsection{A single second order velocity}

The theory \eqref{eq:LagRLT2Vels} depends on two second order velocities $(v_\circ,v_{2\circ})$ and $(v_\perp,v_{2\perp})$ or equivalently $(v_+,v_{2+})$ and $(v_-,v_{2-})$. Therefore, the second order Lagrangian depends on a $n$-dimensional position variable and a $n(n+3)$ dimensional velocity variable, whereas the first order Lagrangian depends on a $n$-dimensional position variable and a $n$-dimensional. Thus, we have introduced $n(n+2)$ degrees of freedom compared to the first order case.
\par 

We will now fix some of the new degrees of freedom by imposing a relation between $v_{2+}$ and $v_{2-}$. 
In order to do this, we recall eqs.~\eqref{eq:DifferentialImposed} and \eqref{eq:Vel2B}
\begin{align*}
	\rd_\pm X^\mu(\tau) 
	&= b_\pm^\mu \, d\tau^{1/2} + v_\pm^\mu \, d\tau + o(d\tau) \,, \nonumber\\
	v_{2\pm}^{\mu\nu}(x,\tau)
	&= \E\left[ \pm b_\pm^\mu \, b_\pm^\nu \, \Big| \, X(\tau)=x \right] .
\end{align*}
Now note that, if $X\in\mathcal{C}^{1/2}(\mathcal{T},\M)$, one has $b_+=b_-$, such that
\begin{equation}\label{eq:Vel}
	v_{2+} = - v_{2-} \, .
\end{equation}
As discussed in section \ref{sec:PIProp},  $X\in\mathcal{C}^{\alpha}(\mathcal{T},\M)$ only for $\alpha<1/2$, such that $b_+\neq b_-$. Nevertheless, in order to reduce the number of degrees of freedom, one may still impose a weaker assumption such as \eqref{eq:Vel}. We will impose a slightly more general condition that there exists only one real second order velocity field $v_2$, such that
\begin{equation}\label{eq:Vel2}
	v_{2\pm} = \alpha_\pm \, v_2 \qquad \textrm{with} \quad \alpha_\pm\in\mathbb{C}\, 
\end{equation}
or equivalently
\begin{equation}\label{eq:v2fixing}
	v_{2^\circ_\perp} = \alpha_{^\circ_\perp} \, v_2 \qquad \textrm{with} \quad \alpha_{^\circ_\perp} = \frac{\alpha_+\pm\alpha_-}{2} \, . 
\end{equation}
Then, we are left with $n(n+5)/2$ degrees of freedom for the velocity variable. 
\par 

After imposing eq.~\eqref{eq:v2fixing}, one obtains the Lagrangian
\begin{align}\label{eq:Lagrangian}
	L_{\rm RLT} 
	&=
	\frac{g_{\mu\nu}}{2\, \varepsilon} \left(
		\hat{v}_\circ^\mu \hat{v}_\circ^\nu 
		+ \beta_2^2 \, \hat{v}_\perp^\mu \hat{v}_\perp^\nu
		+ 2 \, \beta_1 \, \hat{v}_\circ^\mu \hat{v}_\perp^\nu
	\right) 
	+ \frac{\hat{G}_{|\mu\nu|\rho\sigma|}}{2 \, \varepsilon} \left(
		\alpha_\circ^2
		+ \alpha_\perp^2 \beta_2^2
		+ 2 \, \alpha_\circ \alpha_\perp \beta_1
	\right)
	\hat{v}_{2}^{\mu\nu} \hat{v}_{2}^{\rho\sigma}
	\nonumber\\
	&\quad
	+ q \, A_\mu \left(
		v_\circ^\mu
		+ \beta_1 \, v_\perp^\mu
	\right)
	+ \frac{q \left( \alpha_\circ + \alpha_\perp \beta_1 \right)}{2} \, \nabla_\nu A_\mu \hat{v}_2^{\mu\nu}
	- \frac{\varepsilon \, m^2}{2} \, ,
\end{align}
The momenta are given by
\begin{align}
	\hat{p}^{\circ}_\mu 
	&= \frac{g_{\mu\nu}}{\varepsilon} \left(
		\hat{v}_\circ^\nu 
		+ \beta_1 \, \hat{v}_\perp^\nu 
	\right)
	+ q \, A_\mu 
	\, ,\\
	\hat{p}^{\perp}_{\mu} 
	&= \frac{g_{\mu\nu}}{\varepsilon} \left(
		\beta_1 \, \hat{v}_\circ^\mu 
		+ \beta_2^2 \, \hat{v}_\perp^\nu
	\right)
	+ \beta_1 \, q \, A_\mu
	, \\
	\hat{p}^{(2)}_{\mu\nu} 
	&= \frac{\hat{G}_{|\mu\nu|\rho\sigma|}}{\varepsilon} \left(
		\alpha_\circ^2
		+ \alpha_\perp^2 \beta_2^2
		+ 2 \,\alpha_\circ \alpha_\perp \beta_1
	\right)
	\hat{v}_{2}^{\rho\sigma} 
	+ \frac{q \left( 
		\alpha_\circ + \alpha_\perp \, \beta_1
		\right)}{2} \, \nabla_\nu A_\mu
	\, .
\end{align}
Furthermore, the Hamiltonian becomes
\begin{align}\label{eq:Hamiltonian}
	H_{\rm RLT}
	&=
	\frac{\varepsilon \, g^{\mu\nu}}{2\left( \beta_2^2 - \beta_1^2\right)} \Big[
		\beta_2^2 \left(\hat{p}^\circ_\mu - q\, A_\mu \right) \left(\hat{p}^\circ_\nu - q\, A_\nu \right)
		+ \left(\hat{p}^\perp_\mu - \beta_1 \, q\, A_\mu \right) \left(\hat{p}^\perp_\nu - \beta_1 \, q\, A_\nu \right)
	\nonumber\\
	&\qquad \qquad \qquad
		-2\, \beta_1 \left(\hat{p}^\circ_\mu - q\, A_\mu \right) \left(\hat{p}^\perp_\nu - \beta_1 \, q\, A_\nu \right)
	\Big]
		+ \frac{\varepsilon \, m^2}{2} 
	\nonumber\\
	&\quad
	+ \frac{
		\varepsilon \, \hat{G}^{|\mu\nu|\rho\sigma|} \left[ 
			2 \, \hat{p}^{(2)}_{\mu\nu} - q \left(\alpha_\circ + \alpha_\perp \beta_1\right) \nabla_\nu A_\mu 
		\right] \left[ 
			2\, \hat{p}^{(2)}_{\rho\sigma} - q \left(\alpha_\circ + \alpha_\perp \beta_1\right) \nabla_\sigma A_\rho 
		\right] 
	}{ 
		8 \left( \alpha_\circ^2 + \alpha_\perp^2 \beta_2^2 + 2 \, \alpha_\circ \alpha_\perp \beta_1 \right)	
	} 
	\, .
\end{align}

\subsection{Discussion of the theory}
The theory developed in this section has various interesting features, which will be discussed in this subsection. In order to discuss these features, we will first perform a change of variables to a new basis $(\mathfrak{v},\mathfrak{u},\mathfrak{v}_2)$. In particular, we introduce the velocities
\begin{align}
	\mathfrak{v} &= v_\circ + \beta_1 \, v_\perp \, ,\\
	\mathfrak{u} &= \sqrt{\beta_2^2 - \beta_1^2} \, v_\perp \, ,\\
	\mathfrak{v}_2 &= \left( \alpha_\circ + \alpha_\perp \beta_1 \right) v_2 \, ,
\end{align}
such that the Lagrangian is given by
\begin{align}\label{eq:Lagrangian2}
	L_{\rm RLT} 
	&=
	\frac{g_{\mu\nu}}{2\, \varepsilon} \left(
	\hat{\mathfrak{v}}^\mu \hat{\mathfrak{v}}^\nu
	+ \hat{\mathfrak{u}}^\mu \hat{\mathfrak{u}}^\nu
	\right) 
	+ \frac{\gamma^2}{2\, \varepsilon} \, \hat{G}_{|\mu\nu|\rho\sigma|}
	\hat{\mathfrak{v}}_{2}^{\mu\nu} \hat{\mathfrak{v}}_{2}^{\rho\sigma}
	+ q \, A_\mu \hat{\mathfrak{v}}^\mu
	+ \frac{q}{2} \, \nabla_\nu A_\mu \hat{\mathfrak{v}}_2^{\mu\nu}
	- \frac{\varepsilon \, m^2}{2} \, ,
\end{align}
where
\begin{equation}
	\gamma^2 = \frac{ \alpha_\circ^2 + 2 \, \alpha_\circ \alpha_\perp \beta_1 + \alpha_\perp^2 \beta_2^2}{\alpha_\circ^2 + 2 \, \alpha_\circ \alpha_\perp \beta_1 + \alpha_\perp^2 \beta_1^2 } \, .
\end{equation}
The momenta are now given by
\begin{alignat}{2}
	\hat{\mathfrak{p}}_\mu 
	&= \frac{g_{\mu\nu}}{\varepsilon} \, \hat{\mathfrak{v}}^\nu + q \, A_\mu 
	&&= \hat{p}^\circ_\mu \, ,\\
	\hat{\mathfrak{q}}_\mu 
	&= \frac{g_{\mu\nu}}{\varepsilon} \, \hat{\mathfrak{u}}^\nu \, 
	&&= \frac{\hat{p}^\perp_\mu - \beta_1 \hat{p}^\circ_\mu}{\sqrt{\beta_2^2 - \beta_1^2}}\, ,\\
	\hat{\mathfrak{p}}^{(2)}_{\mu\nu} 
	&= \gamma^2 \, \frac{\hat{G}_{|\mu\nu|\rho\sigma|}}{\varepsilon} 
	\hat{\mathfrak{v}}_{2}^{\rho\sigma} 
	+ \frac{q}{2} \, \nabla_\nu A_\mu
	&&= \frac{\hat{p}^{(2)}_{\mu\nu}}{\alpha_\circ + \alpha_\perp \beta_1} \, ,
\end{alignat}
and the Hamiltonian can be rewritten as
\begin{align}\label{eq:Hamiltonian2}
	H_{\rm RLT}
	&=
	\frac{\varepsilon}{2} \left[ 
	g^{\mu\nu} \left(\hat{\mathfrak{p}}_\mu - q\, A_\mu \right) \left(\hat{\mathfrak{p}}_\nu - q\, A_\nu \right)
	+ g^{\mu\nu} \hat{\mathfrak{q}}_\mu \hat{\mathfrak{q}}_\nu
	+ \frac{ \hat{G}^{|\mu\nu|\rho\sigma|} }{ \gamma^2	} \left( 
	\hat{\mathfrak{p}}^{(2)}_{\mu\nu} - \frac{ q }{2} \, \nabla_\nu A_\mu 
	\right) \left( 
	\hat{\mathfrak{p}}^{(2)}_{\rho\sigma} - \frac{ q }{2} \, \nabla_\sigma A_\rho 
	\right)
	+ m^2 
	\right].
\end{align}

\subsubsection{Energy-momentum relation}
The energy-momentum relation, which is the Euler-Lagrange equation for the auxiliary variable $\varepsilon$, can be read off from the Hamiltonian by setting $H=0$. It follows, using the Hamiltonian \eqref{eq:Hamiltonian2}, that the energy-momentum relation in second order geometry is modified to
\begin{align}\label{eq:EMEq}
	g^{\mu\nu} \left( \hat{\mathfrak{p}}_\mu - q \, A_\mu \right) \left( \hat{\mathfrak{p}}_\nu - q \,  A_\nu \right) 
	+ g^{\mu\nu} \hat{\mathfrak{q}}_\mu \hat{\mathfrak{q}}_\nu 
	+ \frac{ \hat{G}^{|\mu\nu|\rho\sigma|}}{ \gamma^2} \left( 
	\hat{\mathfrak{p}}^{(2)}_{\mu\nu} - \frac{ q }{2} \, \nabla_\nu A_\mu 
	\right) \left( 
	\hat{\mathfrak{p}}^{(2)}_{\rho\sigma} - \frac{ q }{2} \, \nabla_\sigma A_\rho 
	\right)
	= - m^2  \, .
\end{align}
Then, if one imposes the first order energy-momentum relation
\begin{equation}\label{eq:EMEq1}
	g^{\mu\nu} 
	\left(\hat{p}^\circ_\mu - q \, A_\mu\right) \left(\hat{p}^\circ_\nu - q \, A_\nu\right) 
	= - m^2 \, ,
\end{equation}
one is left with the additional equation for the second order parts
\begin{equation}\label{eq:EMEq2}
	g^{\mu\nu} \hat{\mathfrak{q}}_\mu \hat{\mathfrak{q}}_\nu 
	+ \frac{ \hat{G}^{|\mu\nu|\rho\sigma|}}{ \gamma^2} \left( 
	\hat{\mathfrak{p}}^{(2)}_{\mu\nu} - \frac{ q }{2} \, \nabla_\nu A_\mu 
	\right) \left( 
	\hat{\mathfrak{p}}^{(2)}_{\rho\sigma} - \frac{ q }{2} \, \nabla_\sigma A_\rho 
	\right) 
	= 0 \, .
\end{equation}

\subsubsection{Ostragradski's instability}
The Hamiltonian \eqref{eq:Hamiltonian2} is quadratic in all its momenta. Thus, for a positive definite metric, the Hamiltonian can be bounded from below, and one expects the theory to be free of instabilities. This feature may come as some surprise due to Ostragradski's theorem, cf. e.g. \cite{Woodard:2015zca} for a review. Indeed, second order geometry makes explicit use of second order derivatives, whereas, according to Ostragradski's theorem, higher derivative theories should suffer from ghost-like instabilities. In particular, the addition of an acceleration term $d^2 x/d\tau^2$ to the Lagrangian is known to be a source of the Ostrgradski instability. Hence, one may expect that the addition of the acceleration-like velocities $v_\perp$ and $v_2$ to the Lagrangian will generate such instabilities.
However, there is an important difference between $v_\perp$ and an acceleration $\ddot{X}$. Indeed, in classical theories acceleration is evaluated at order $d\tau^2$:
\begin{equation}
	\ddot{X} = \frac{\rd^2 X}{d\tau^2} \qquad {\rm and} \qquad \dot{X}^2 =  \frac{\rd X}{d\tau} \frac{\rd X}{d\tau}\, ,
\end{equation}
whereas the velocities $v_\perp$ and $v_2$ are evaluated at order $d\tau$:
\begin{equation}
	v_\perp = \frac{1}{2} \, \frac{\rd^2 X}{d\tau} \qquad {\rm and} \qquad v_2 =  \frac{\rd X \rd X}{d\tau}\, .
\end{equation}
This order mixing in $d\tau$ is an essential property of both quantum and stochastic theories, which is built into second order geometry, and allows to evade Ostragradski's theorem.\footnote{A similar argument has been presented in Ref.~\cite{Grudka:2024llq}.} In other words, the proof of Ostragradski's theorem relies on using first order geometry and does not generalize to second order geometry. 
Therefore, despite being second order in derivatives, the theory is free of Ostragradski ghosts. 

\subsubsection{Degrees of freedom}
The second order theory \eqref{eq:Lagrangian} is constructed on a real $n$-dimensional  manifold, but it has a real $n(n+5)/2$-dimensional tangent space. Let us denote the tangent space at a point $x\in\M$ by $\tilde{T}_{2x}\M$. The elements of this tangent space can be represented by the tuple $(v_\circ,v_\perp,v_2) \in \tilde{T}_{2x}\M$. Thus, in comparison to the ordinary second order tangent space \eqref{eq:TangentSpaceDecomp}, it contains an additional part due to the presence of two velocity fields as discussed in section \ref{sec:2Vels}. The presence of these three parts allows to decompose the vector space as
\begin{equation}\label{eq:TangSpaceDecomp}
	\tilde{T}_x\M \cong T_x\M \oplus T_x\M \oplus {\rm Sym}(T_x\M \otimes T_x\M) \, .
\end{equation}
\par 

The signature of the tangent spaces is set by the second order metric $G$. As shown in section \ref{sec:Metric}, it depends on the signature of the first order metric $g$ and the choice of parameter $c_2\in\R$. In particular, for a non-relativistic theory on a $3$-dimensional Riemannian manifold, the tangent space is $12$-dimensional with strictly positive signature $(12,0)$, if $c_2>0$, whereas it has a Lorentzian signature $(11,1)$, if $c_2<0$. Similarly, for a  relativistic theory on a $4$-dimensional Lorentzian manifold, the tangent space is $18$-dimensional with signature $(13,5)$ for $c_2>0$ and $(12,6)$ for $c_2<0$.
\par 

The appearance of a large number of negative eigenvalues in the metric may be considered problematic, as it generates ghost-like instabilities in the Hamiltonian. More precisely, due to the negative eigenvalues the Hamiltonian \eqref{eq:Hamiltonian2} is no longer bounded from below. Thus, although being free of the Ostragradski instability, the theory may have other instabilities. 
\par

A possible solution to this problem is that the degrees of freedom associated to negative eigenvalues are not physical, and that the physical degrees of freedom span a subspace of the  tangent space with positive definite signature.
In fact, as shown by eq.~\eqref{eq:conditionFeynman} in quantum theories the second order velocity field is proportional to the metric. More precisely, eq.~\eqref{eq:conditionFeynman} suggests to fix\footnote{The factor $\ri$ that is present in  is eq.~\eqref{eq:conditionFeynman} can be introduced by $\alpha_{1,2,3}$ and/or $\alpha_\pm$.}
\begin{equation}
	v_2^{\mu\nu}(x,\tau) = \hbar \, \varepsilon(\tau) \, g^{\mu\nu}(x) \, .
\end{equation}
As is the case for the condition \eqref{eq:Epsilonconditon}, this relation is a constraint that  must be imposed in the equations of motion, but not at the level of the Lagrangian. The fact that the second order velocity is proportional to the metric implies that it does not have $n(n+1)/2$ physical degrees of freedom, since for the metric only $n(n-1)/2$ degrees of freedom are physical, while the other $n$ are redundant due to the Bianchi identities. As a consequence, in the $3$-dimensional non-relativistic theory, one is left with a non-trivial subspace of $\tilde{T}_x\M$, which is $9$-dimensional and has a strictly positive signature, whereas in the $4$-dimensional relativistic theory, one is left with a non-trivial subspace of $\tilde{T}_x\M$, which is $14$-dimensional with signature $(12,2)$.
\par 

The relativistic theory is further constrained by the energy-momentum relation \eqref{eq:EMEq}, which reduces the physical tangent space to be $13$-dimensional with signature $(12,1)$. In order to remove the last negative eigenvalue, one expects the existence of at least one more constraint on the theory. This final constraint could be obtained by imposing the first and second order energy-momentum relation separately, instead of imposing the general energy-momentum relation \eqref{eq:EMEq}. In this case the energy-momentum relation imposes two constraints \eqref{eq:EMEq1} and \eqref{eq:EMEq2}, such that the physical tangent space becomes $12$-dimensional with strictly positive signature.

\subsubsection{Time-reversal symmetry}
The theory defined by the Lagrangian \eqref{eq:Lagrangian0} depends on five parameters $\alpha_1,\alpha_2,\alpha_3,\alpha_+,\alpha_-\in\mathbb{C}$. However, since the theory can be rewritten as \eqref{eq:Lagrangian} or \eqref{eq:Hamiltonian}, it only depends on the four $\beta_1,\beta_2,\alpha_+,\alpha_-\in\mathbb{C}$, where $\beta_{1,2}$ are defined in terms of $\alpha_{1,2,3}$, cf. eq.~\eqref{eq:betadef}. Thus, we find that at most four of these parameters may be physical. The fifth parameter is associated to a gauge redundancy, as it only affects the parameterization of the theory. 
\par 

These remaining parameters may be fixed by theoretical and/or experimental considerations.
An example of such a constraint could come from time-reversal symmetry. Even though time-reversal symmetry can be violated in nature and is replaced by CPT-symmetry, one may study bounds on the parameters set by the time-reversal symmetry. Under a time-reversal symmetry, the role of the left and right limits changes, such that $v_+\leftrightarrow v_-$. When extended to the second order vector, this transformation acts as $(v_+,v_{2+})\leftrightarrow(v_-,-v_{2-})$ due to the $-$ sign in the definition of $v_{2-}$. In the $(v_\circ,v_\perp)$ basis this corresponds to a exchange of the form $(v_\circ,v_{2\circ})\leftrightarrow (v_\circ,-v_{2\circ}) $ and $(v_\perp,v_{2\perp})\leftrightarrow (-v_\perp,-v_{2\perp}) $. By considering the Lagrangian \eqref{eq:Lagrangian}, we find that the theory is invariant under this symmetry, if
\begin{align}
	\alpha_\circ &= - \alpha_\perp \, \beta_1 \, ,\\
	\beta_1 &= 0 \, ,
\end{align}
which implies
\begin{equation}
	\beta_2^2 = - \alpha_1 \, \alpha_2 \, .
\end{equation}
The Lagrangian is then given by
\begin{align}
	L_{\rm RLT} 
	&=
	\frac{g_{\mu\nu}}{2\, \varepsilon} \left(
	\hat{v}_\circ^\mu \hat{v}_\circ^\nu 
	- \alpha_1 \alpha_2 \, \hat{v}_\perp^\mu \hat{v}_\perp^\nu
	\right) 
	- \frac{ \alpha_1 \alpha_2 \alpha_\perp^2}{2 \, \varepsilon} \, \hat{G}_{|\mu\nu|\rho\sigma|}
	\hat{v}_{2}^{\mu\nu} \hat{v}_{2}^{\rho\sigma}
	+ q \, A_\mu v_\circ^\mu
	- \frac{\varepsilon \, m^2}{2} \, ,
\end{align}
such that one is left with the free parameters $\alpha_1\alpha_2$ and $\alpha_\perp^2$.

\section{Generalizations}\label{sec:Generalizations}
In the previous section \ref{eq:PhysTheory}, we have focused on a physical theory that contains just enough freedom to describe a coupling of quantum theory to gravity. However, throughout the work we made several assumptions that simplified the theory and allowed to arrive at this minimal framework. In this section, we will review these assumptions and discuss what potential generalizations would look like.
\par

A first generalization can be achieved by introducing torsion or metric incompatibility. These generalizations are often studied in first order geometry. As pointed out in section \ref{sec:Metric}, introducing torsion or metric incompatibility in first order geometry will also change the second order data, such as the second order metric. 
\par 

More generally, one may also drop the minimal assumption that the second order metric is determined in terms of first order data. In this case, one can introduce new fields in the second order metric. An example would be to introduce an anti-symmetric field $B_{\mu\nu}$ in the metric on the entire tangent space \eqref{eq:TangSpaceDecomp} that explicitly couples the $v_\circ$ sector to the $v_\perp$ sector.\footnote{In the current minimal formulation, there is also a coupling between these sectors, but it is parameterized by $\beta_1\in\mathbb{C}$ instead of a tensor field $B_{\mu\nu}$.} Such a generalization bears similarities with generalized geometry \cite{Hitchin:2003cxu,Gualtieri:2003dx,Gualtieri:2007bq}, which has found applications in string theory through double field theory \cite{Siegel:1993th,Hull:2009mi,Hull:2009zb,Hohm:2010jy,Hohm:2010pp,Hohm:2013bwa}.
\par 

A third generalization that one could consider is the violation of eq.~\eqref{eq:Vel2}. In this case $v_{2+}$ and $v_{2-}$ remain independent fields, which will introduce a further $n(n+1)/2$ degrees of freedom.
\par 

Another generalization would be the study of non-symmetric second order vector fields. In this case the degrees of freedom of $v_{2}$ are extended from $n(n+1)/2$ to $n^2$. Since the symmetry of $v_2$ is inherited from the commutativity of partial derivatives, this extension necessitates the introduction of new derivative operators that are non-commutative. As in gauge theories, this can be realized by introducing derivative operators 
\begin{equation}
	D_\mu = \p_\mu + \chi_\mu \, ,
\end{equation}
which satisfy the non-commutativity
\begin{equation}
	[D_\mu,D_\nu] 
	= \p_\mu \chi_\nu - \p_\nu \chi_\mu  + [\chi_\mu, \chi_\nu] \, .
\end{equation}
From eq.~\eqref{eq:Vel2B} it follows that an asymmetry of $v_2$ implies a non-commutativity for the objects $b_\pm$. Therefore, for non-symmetric $v_{2}$, one requires
\begin{equation}
	[b_\pm^\mu, b_\pm^\nu ] = \dot{B}_\pm^{\mu\nu}(X,\tau) \neq 0 \, ,
\end{equation}
which implies
\begin{equation}
	\int [\rd_\pm X^\mu, \rd_\pm X^\nu ] = \int \dot{B}_\pm^{\mu\nu}(X,\tau) \, d\tau \neq 0\, .
\end{equation}
This suggests a non-commutativity of the form
\begin{equation}
	[x^\mu, x^\nu] = B^{\mu\nu}(x) \, .
\end{equation}
Therefore, one expects that this type of spacetime non-commutativity can be encoded using non-symmetric second order vectors.
\par 

As a final generalization one can study theories with less regularity, which requires higher order geometry. In particular, one can consider paths $X\in C^{1/k}(\mathcal{T},\M)$ with $k\in\mathbb{N}$, which requires $k$th-order geometry, as developed in Ref.~\cite{Bies:2023zvs,Bies:202406,Bies:2024gjn}. One may even consider the limit $k \rightarrow \infty$, which corresponds to a path integral that sums over discrete paths. As shown in \cite{Arzano:2024apl}, this scenario can be associated with another type of spacetime non-commutativity, namely one that involves non-commutativity relations of the form
\begin{equation}
	[x^\mu, x^\nu] = C^{\mu\nu}_{\;\;\rho}(x)\, x^\rho \, .
\end{equation}

\section{Conclusions \& outlook}\label{sec:Conclusion}

We have argued that the study of quantum theories subjected to gravity requires extensions beyond first order Riemannian geometry. Inspired by the path integral formulation, we provided a concrete proposal of such an extension in the form of second order geometry \cite{Meyer:1981,Schwartz:1984,Emery:1989,Huang:2022}. We then provided a basic introduction to this framework, and discussed how physical theories can be constructed within second order geometry.
\par

Although our discussion of physical theories was restricted to the study of a single scalar quantum particle on a curved space(time), there are some lessons that can be drawn from this analysis that will be relevant for the development of field theories within second order geometry. In particular, the presence of higher derivative terms implies that the correct configuration space for field theories is the second order jet bundle instead of the first order jet bundle that is considered in classical field theory. Thus, a field theory Lagrangian for a scalar field will be given by $L(\phi,\nabla\phi,\nabla\nabla\phi)$, which should appear in the path integral measure instead of $L(\phi,\nabla\phi)$. Another implication for field theory is that the generalization of the energy-momentum relation \eqref{eq:EMEq} implies a generalization of the dispersion relations for field theory. This generalization of the dispersion relations may necessitate the introduction of second order Fourier analysis and the study of second order Hilbert spaces, as discussed in \cite{Tao:2016}.
\par

The motivation for studying second order geometry was provided by the construction of a manifestly covariant path integral on curved spacetimes, and thus primarily for constructing quantum theories on a curved spacetime. Nevertheless, the framework could also be useful for the study of quantum gravity. The study of field theories and, in particular, the spin-2 metric field within second order geometry is beyond the scope of this paper. However, the fact that second order geometry introduces terms that are quadratic in derivatives, while avoiding the Ostragradski instability, suggests that quadratic gravity \cite{Stelle:1976gc,Stelle:1977ry} may be both renormalizable and ghost-free, when interpreted within second order geometry.
\par

Coincidentally, we also recovered some elementary ideas encountered in quantum gravity theories. In particular, we pointed out in section \ref{sec:PIProp} that, in the path integral formulation of quantum mechanics, particle trajectories span a 2-dimensional fractal surface, which could be thought of as the worldsheet generated by a propagating excited string. Furthermore, in the construction of the second order metric in section \ref{sec:Metric}, we found that such a metric contains two parts: the first order part that defines a line element and a second part that defines an area element. The latter appears in various approaches to quantum gravity, cf. e.g. \cite{Schuller:2005yt,Schuller:2005ru,Punzi:2006hy,Punzi:2006nx,Dittrich:2023ava,Borissova:2023yxs}. Moreover, we saw that the construction of the second order part forces us to introduce a small fundamental length/area scale. This scale can be used as a natural cutoff scale for the theory and can be compared to other cutoff scales encountered in various quantum gravity theories, such as the string scale or a minimal length scale associated to the discretization of geometry.
\par

In addition, we discussed possible relations with non-commutative geometry, which is regarded to be a candidate for a `quantum geometry' \cite{Beggs:2020}, and is known to have connections to quantum gravity frameworks such as string theory \cite{Sheikh-Jabbari:1997qke,Ardalan:1998ce,Connes:1997cr,Seiberg:1999vs}. In particular, we found that, depending on the type of non-commutativity that is being studied, non-commutativty can either arise as the infinite order limit of higher order geometry or due to the presence of asymmetric second order vector fields in second order geometry.
\par 

Similarly, we also pointed out possible relations to generalized geometry  \cite{Hitchin:2003cxu,Gualtieri:2003dx,Gualtieri:2007bq,Siegel:1993th,Hull:2009mi,Hull:2009zb,Hohm:2010jy,Hohm:2010pp,Hohm:2013bwa}. In generalized geometry, one studies tangent spaces that can be decomposed as a direct sum of the tangent and cotangent space: $\tilde{T}_x\M\cong T_x\M \oplus T^\ast_x\M$. In this context, generalized geometry might be regarded as a milder modification of Riemannian geometry, where the second order part $v_2$ is neglected and only the extension to two velocity fields $(v_\circ,v_\perp)$ is made. We note that second order geometry introduces a coupling between the second order vectors $(v_2)$ and the sector $(v_\circ,v_\perp)$, but it does not necessarily introduce a coupling between $v_\circ$ and $v_\perp$. On the other hand, in generalized geometry there is no $(v_2)$ sector, but there is a coupling between $v_\circ$ and $v_\perp$ introduced by the $B$-field. From the second order geometry perspective this coupling could be introduced by hand as discussed in previous section \ref{sec:Generalizations}, but, in principle, it could also be obtained as an effective coupling obtained by integrating out the second order sector associated to $v_2$.
\par

As in generalized geometry, the construction of a Lie bracket in second order geometry requires a modification of its definition, since the standard Lie bracket of two second order vector fields does not close on itself. Instead of modifying the bracket, one could also define the bracket on a restricted class of second order vectors  \cite{Kuipers:2021jlh,Huang:2022}, which provides a first attempt in constructing second order Killing vectors \cite{Kuipers:2022wpy}. Alternatively, one could define the Lie bracket only within infinite order geometry \cite{Bies:2023zvs}. The latter suggests that the study of $C^{1/2}$ paths forces one to also consider $C^{1/k}$ paths for any $k\in\mathbb{N}$. This corresponds to the study of higher dimensional branes, since the upper bound on the Hausdorff dimension of quantum paths, as discussed in section \ref{sec:PIProp}, is set by the H\"older regularity as ${\rm dim_H}\leq k$ \cite{Morters:2010}.
\par

We emphasize that at this point these relations between higher order geometry and quantum gravity or quantum geometry are rather superficial. Nevertheless, they could serve as a starting point for the study of possible connections between the various frameworks, which might aid towards a further development of quantum geometry and a theory of quantum gravity.
\par 

Another potential avenue for further development of the framework is the study of the unification of forces within the higher order geometry framework, as done in Ref.~\cite{Bies:2024gjn}. It is noteworthy that the unification of forces has also been studied in the literature \cite{Percacci:1984ai,Percacci:1990wy,Nesti:2009kk,Chamseddine:2010rv,Chamseddine:2013hwa,Chamseddine:2016pkx,Krasnov:2017epi} by considering higher dimensional tangent spaces attached to a $4$-dimensional manifold. In particular, in recent works it was proposed that this can be realized for an $18$-dimensional tangent space \cite{Konitopoulos:2023wst,Roumelioti:2024lvn}. It would be interesting to investigate, whether the $18$-dimensional tangent spaces obtained in second order geometry can be made compatible with such a scenario.
\par 

In conclusion, second order geometry and its generalization to higher order geometry can elucidate the interplay between quantum theory and gravity, both by itself and in relation to other theories.

\section*{Acknowledgements}
The work of F.K. is supported by a postdoctoral fellowship of the Alexander von Humboldt foundation and a fellowship supplement of the Carl Friedrich von Siemens foundation.

\appendix

\section{Martingales and position operators}\label{Sec:Martingales}
In this appendix, we provide some further mathematical background for the objects $b_\pm$ introduced in section \ref{sec:Differentials}. We start by recalling the formal equation \eqref{eq:DifferentialImposed}:
\begin{equation*}
	\rd_\pm X^\mu(\tau) = b_\pm^\mu(X(\tau),\tau) \, d\tau^{1/2} + v_\pm^\mu(X(\tau),\tau) \, d\tau + o(d\tau) \, .
\end{equation*}
This expression can be integrated in a given coordinate chart\footnote{The definition of a covariant version of such an integral on manifolds is the topic of second order geometry and requires the second order forms as discussed in section \ref{eq:2ndGeom} \cite{Emery:1989}.}, which yields
\begin{equation}
	\int \rd_\pm X^\mu(\tau) 
	= \int b_\pm^\mu(X(s),s) \, ds^{1/2} 
	+ \int v_\pm^\mu(X(s),s) \, ds  \, .
\end{equation}
More generally, one may consider the integral over some integrable function $f$:
\begin{equation}\label{eq:IntExp}
	\int f(X(\tau),\tau)\, \rd_\pm X^\mu(\tau) 
	= \int f(X(s),s) \, b_\pm^\mu(X(s),s) \, ds^{1/2} 
	+ \int f(X(s),s) \, v_\pm^\mu(X(s),s) \, ds  \, .
\end{equation}
The integral in this expression is defined as the limit of a discretization:
\begin{align}\label{eq:LimIntA}
	\int_{\tau_0}^\tau f(X(\tau),\tau) \,\rd_+ X^\mu(\tau)  
	&= 
	\lim_{K\rightarrow \infty} \sum_{k=0}^{K-1}
	f(X(\tau_k),\tau_k) \, \left[  X^\mu(\tau_{k+1}) - X^\mu(\tau_{k}) \right] ,\\
	\int_{\tau_0}^\tau f(X(\tau),\tau) \,\rd_- X^\mu(\tau)  
	&= 
	\lim_{K\rightarrow \infty} \sum_{k=1}^{K}
	f(X(\tau_k),\tau_k) \, \left[  X^\mu(\tau_{k}) - X^\mu(\tau_{k-1}) \right] ,\label{eq:LimIntB}
\end{align}
where $\{\tau_0,\tau_1,...,\tau_K=\tau\}$ partitions the interval $[\tau_0,\tau]$. It immediately follows that
\begin{equation}
	X^\mu(\tau) - X^\mu(\tau_0) = \int_{\tau_0}^\tau \rd_\pm X^\mu(\tau)\, ,
\end{equation}
where the left hand-side is independent of the choice for $\rd_+ X$ or $\rd_-X$. We emphasize, however, that this only holds for $f(X)=1$, while for a generic function $f$, one has
\begin{equation}
	\int_{\tau_0}^\tau f(X(\tau),\tau)\, \rd_+ X^\mu(\tau) \neq \int_{\tau_0}^\tau f(X(\tau),\tau)\, \rd_- X^\mu(\tau) \, .
\end{equation}
These integral expressions can be used to provide meaning to the right hand side of eq.~\eqref{eq:DifferentialImposed}. Indeed, if the limits \eqref{eq:LimIntA} and \eqref{eq:LimIntB} exist and are independent of the chosen partition, the integral \eqref{eq:IntExp} exists. In that case, $b_\pm$ can be defined in terms of its integral action.
\par 

We note that the second integral in eq.~\eqref{eq:IntExp} is a Lebesgue-Stieltjes integral, for which the existence criteria are well-known. Thus, we may focus on the first integral. This integral can be rewritten as
\begin{align}\label{eq:Bmartingale}
	\int_{\tau_0}^\tau f(X(s),x) \, b_\pm^\mu(X(s),s) \, \, ds^{1/2} 
	&= \int_{\tau_0}^\tau f(X(s),x) \, e^\mu_{\; a}(X(s)) \, b_\pm^a(s) \, \, ds^{1/2} \nonumber\\
	&= \int_{\tau_0}^\tau f(X(s),s) \, e^\mu_{\; a}(X(s)) \, \rd_\pm M^a(s)  \, ,
\end{align}
where $e^\mu_{\;a}$ is a soldering form. 
We may now impose $X$ to be a stochastic process. More precisely, we may impose it to be a semi-martingale such that it can be decomposed as $X=C+M$, where $C$ is deterministic and $M$ is a martingale. In that case, this integral is an \textit{It\^o integral}, for which the existence criteria are also well-known, cf. e.g. Ref.~\cite{KaratzasSchreve}. As discussed in section \ref{sec:PIProp}, we expect $M$ to be some continuous Wiener-type process, which implies that it is indeed a martingale.
\par 

Additionally, we remark that, if $M$ is a Wiener-type process, it can be decomposed as \cite{Biane:2010}
\begin{equation}
	M(\tau) = a(\tau) + a^\ast(\tau) \, ,
\end{equation}
where $a,a^\ast$ are creation and annihilation operators acting on the Fock space $\mathcal{F}(L^2(\R_+)\otimes \mathbb{C}^n)$. Hence,
\begin{equation}
	M(\tau) = \int_0^\tau \rd_0 M(s)
\end{equation}
is a position operator, that can be decomposed into creation and annihilation operators. This suggests a relation between the objects
\begin{equation}
	\int b_\pm \, ds^{1/2} 
\end{equation}
and the ladder operators in quantum mechanics.

\section{Higher order differentials}\label{ap:higherOrder}

In this appendix, we generalize the results of sections \ref{sec:Differentials} and \ref{sec:DiffOperators} to higher order.

\subsection{Differentials}\label{Ap:HighOrderDiff}
In this section, we calculate the differentials $\rd_a^kX$ for any $k\in\mathbb{N}$ in the special cases $a\in\{-1,0,1\}$. For $a=0$, we find
\begin{align}\label{Calc:rd0k}
	\rd_0^k X(\tau) 
	&=
	\sum_{m=0}^k (-1)^{m} {k \choose m} \, X\Big[ \tau + \frac{k-2\,m}{2}\, d\tau\Big] 
	\nonumber\\
	&=
	\sum_{m=0}^k (-1)^{m} {k \choose m} \,\Big[  X\Big(  \tau + \frac{k-2\,m}{2}\, d\tau\Big) - X(\tau) \Big]
	\nonumber\\
	&=
	\sum_{m=0}^{\lfloor k/2\rfloor} (-1)^m {k \choose m} \left\{
	\sqrt{\frac{k-2\,m}{2}} \left[ b_+ - (-1)^k \, b_- \right] d\tau^{1/2}
	+ \frac{k-2\,m}{2} \left[ v_+ - (-1)^k \, v_- \right] d\tau 
	\right\} ,
\end{align}
such that
\begin{alignat}{2}
	\lim_{d\tau\rightarrow0} \E\left[\frac{\rd_0 X(\tau)}{d\tau} \, \Big| \, X(\tau) = x \right]
	&=
	v_\circ \, ;
	\nonumber\\
	\lim_{d\tau\rightarrow0} \E\left[\frac{\rd_0^{2k} X(\tau)}{2\,d\tau} \, \Big| \, X(\tau) = x \right]
	&=
	(-1)^{k+1} {2\, k-2 \choose k - 1 } \, v_\perp \, ,
	\qquad && \forall\, k\geq1\, ;
	\nonumber\\
	\lim_{d\tau\rightarrow0} \E\left[\frac{\rd_0^{2k+1} X(\tau)}{2\,d\tau} \, \Big| \, X(\tau) = x \right]
	&=
	0 \, ,
	\qquad && \forall\, k\geq1 \, .
\end{alignat}
Moreover, for $a\in\{-1,1\}$, we obtain
\begin{align}
	\rd_+^k\rd_-^l X 
	&= \sum_{m=0}^{k+l} (-1)^m {k+l\choose m} X(\tau+(k-m)\,d\tau)
	\nonumber\\
	&= \sum_{m=0}^{k+l} (-1)^m {k+l\choose m} \Big[ X(\tau+(k-m)\,d\tau) - X(\tau)\Big]
	\nonumber\\
	&= \sum_{m=0}^{k-1} (-1)^m {k+l\choose m} \left[ \sqrt{k-m}\, b_+ \, d\tau^{1/2} + (k-m) \, v_+ \, d\tau \right]
	\nonumber\\
	&\quad
	- \sum_{m=k+1}^{k+l} (-1)^m {k+l\choose m} \left[ \sqrt{m-k} \, b_- \, d\tau^{1/2} + (m-k) \, v_- \, d\tau \right] .
\end{align}
Then, calculating the limits yields
\begin{alignat}{2}
	\lim_{d\tau\rightarrow0} \E\left[\frac{\rd_+^k X(\tau)}{2\,d\tau} \, \Big| \, X(\tau) = x \right]
	&=
	0 \,, \qquad &&\forall \, k\geq2 \, ;
	\nonumber\\
	\lim_{d\tau\rightarrow0} \E\left[\frac{\rd_-^k X(\tau)}{2\,d\tau} \, \Big| \, X(\tau) = x \right]
	&=
	0 \,, \qquad &&\forall \, k\geq2 \, ;
	\nonumber\\
	\lim_{d\tau\rightarrow0} \E\left[\frac{\rd_+^k\rd_-^l X(\tau)}{2\,d\tau} \, \Big| \, X(\tau) = x \right]
	&=
	(-1)^{k+1} \, {k+l-2 \choose k-1} \, v_\perp \, ,\qquad &&\forall\,k,l\geq1 \, .
\end{alignat}

\subsection{Differential operators}\label{Ap:HighOrderDiffOp}
In this section, we calculate differential operators of the form $\rd_a^kf$ for any $k\in\mathbb{N}$ in the special cases $a\in\{-1,0,1\}$. At first order, we obtain for $a=0$
\begin{align}\label{Calc:d0f}
	\rd_0 f(X)
	&:= f(X(\tau+d\tau/2)) -  f(X(\tau-d\tau/2)) \nonumber\\
	&= [f(X(\tau+d\tau/2)) - f(X(\tau))] + [f(X(\tau)) - f(X(\tau-d\tau/2))] \nonumber\\
	&= \p_\mu f(X) \, \left[X^\mu(\tau+d\tau/2) - X^\mu(\tau-d\tau/2) \right]
	\nonumber\\
	&\quad
	+ \frac{1}{2} \p_\nu \p_\mu f(X) \, \Big\{ 
	\left[X^\mu(\tau+d\tau/2) - X^\mu(\tau) \right] \left[X^\nu(\tau+d\tau/2) - X^\nu(\tau) \right]
	\nonumber\\
	&\qquad\qquad\qquad\quad
	- \left[X^\mu(\tau) - X^\mu(\tau-d\tau/2) \right] \left[X^\nu(\tau) - X^\nu(\tau-d\tau/2) \right] \Big\}
	\nonumber\\
	&=  \p_\mu f(X) \, \rd_0 X^\mu + \frac{1}{4} \, \p_\nu \p_\mu f(X) \left( \rd_+ X^\mu \rd_+X^\nu - \rd_- X^\mu \rd_-X^\nu \right) + o(d\tau) \, . 
\end{align}
We can generalize this to higher orders. For $a=0$, one obtains
\begin{align}\label{Calc:d0kf}
	\rd_0^k f(X) 
	&= \sum_{m=0}^{k} (-1)^m {k\choose m} f\Big[X\Big(\tau + \frac{k-2m}{2}\,d\tau\Big)\Big]
	\nonumber\\
	&= \sum_{m=0}^{k} (-1)^m {k\choose m} \Big\{ f\Big[X\Big(\tau + \frac{k-2m}{2}\,d\tau\Big)\Big] - f[X(\tau)] \Big\}
	\nonumber\\
	&= \sum_{m=0}^{k} (-1)^m {k\choose m} \left\{ \p_\mu f(X)\, \Big[ X^\mu\Big(\tau + \frac{k-2\,m}{2}\,d\tau\Big) - X^\mu(\tau) \Big] 
	\right. \nonumber\\
	&\quad \left.
	+ \frac{1}{2} \, \p_\nu\p_\mu f(X) \, \Big[ X^\mu\Big(\tau + \frac{k-2\,m}{2}\,d\tau\Big) - X^\mu(\tau) \Big]  \, \Big[ X^\nu\Big(\tau + \frac{k-2\,m}{2}\,d\tau\Big) - X^\nu(\tau) \Big]  \right\}
	\nonumber\\
	&= \p_\mu f(X) \, \rd_0^k X^\mu
	+ \frac{1}{2} \, \p_\nu \p_\mu f(X) \left\{ 
	\sum_{m=0}^{\lfloor k/2\rfloor} (-1)^m {k\choose m} \, \frac{k-2\,m}{2} \, \rd_+ X^\mu \rd_+ X^\nu
	\right. \nonumber\\
	&\qquad \left.
	+ \sum_{m=\lceil k/2\rceil}^{k} (-1)^m {k\choose m} \frac{2\,m-k}{2} \, \rd_- X^\mu \rd_- X^\nu
	\right\} + o(d\tau)
	\nonumber\\
	&=
	\p_\mu f(X) \, \rd_0^k X^\mu + \frac{1}{2} \, \p_\nu \p_\mu f(X) 
	\nonumber\\
	&\quad 
	\times
	\begin{cases}
		0 \qquad &{\rm if} \quad k\geq2 \; \textrm{is odd;}\\
		(-1)^{\frac{k}{2}+1} {k -2 \choose k/2 -1} \, (\rd_+X^\mu \rd_+X^\nu + \rd_-X^\mu \rd_-X^\nu)\, , \qquad &{\rm if} \quad k\geq2 \; \textrm{is even.}
	\end{cases}
\end{align}
Furthermore, for $a\in\{-1,1\}$, one obtains
\begin{align}\label{Calc:dkdlf}
	\rd_+^k\rd_-^l f(X) 
	&= \sum_{m=0}^{k+l} (-1)^m {k+l\choose m} f[X(\tau + (k-m)\,d\tau)]
	\nonumber\\
	&= \sum_{m=0}^{k+l} (-1)^m {k+l\choose m} \Big\{ f[X(\tau + (k-m)\,d\tau)] - f[X(\tau)] \Big\}
	\nonumber\\
	&= \sum_{m=0}^{k+l} (-1)^m {k+l\choose m} \Big\{ \p_\mu f(X) \, \Big[ X^\mu(\tau + (k-m)\,d\tau) - X^\mu(\tau) \Big] 
	\nonumber\\
	&\quad
	+ \frac{1}{2} \, \p_\nu\p_\mu f(X) \, \Big[ X^\mu(\tau + (k-m)\,d\tau) - X^\mu(\tau) \Big] \, \Big[ X^\nu(\tau + (k-m)\,d\tau) - X^\nu(\tau) \Big] \Big\}
	\nonumber\\
	&= \p_\mu f(X) \, \rd_+^k\rd_-^l X^\mu
	+ \frac{1}{2} \, \p_\nu \p_\mu f(X) \left\{ 
	\sum_{m=0}^{k-1} (-1)^m {k+l\choose m} (k-m) \, \rd_+ X^\mu \rd_+ X^\nu
	\right. \nonumber\\
	&\qquad \left.
	+ \sum_{m=k+1}^{k+l} (-1)^m {k+l\choose m} (m-k) \, \rd_- X^\mu \rd_- X^\nu
	\right\} + o(d\tau)
	\nonumber\\
	&=
	\p_\mu f(X) \, \rd_+^k\rd_-^l X^\mu + \frac{1}{2} \, \p_\nu \p_\mu f(X) 
	\nonumber\\
	&\quad 
	\times
	\begin{cases}
		0 \qquad &{\rm if} \quad k=0\, , \; l\geq2 \,;\\
		0 \qquad &{\rm if} \quad k\geq2\, ,\; l=0 \,;\\
		(-1)^{k+1} {k+l-2 \choose k-1}(\rd_+X^\mu \rd_+X^\nu + \rd_-X^\mu \rd_-X^\nu) \qquad &{\rm if} \quad k\geq1\, ,\; l\geq1 \,.
	\end{cases}
\end{align}
\par 

Finally, we can evaluate the limits. This yields
\begin{align}
	\lim_{d\tau\rightarrow 0 } \E\left[ \frac{\rd_\pm^k f(X(\tau))}{2\,d\tau} \, \Big| \, X(\tau)=x\right] 
	&=  0 \qquad \forall \, k\geq2 \, . \nonumber\\
	\lim_{d\tau\rightarrow 0 } \E\left[ \frac{\rd_+^k\rd_-^l f(X(\tau))}{2\,d\tau} \, \Big| \, X(\tau)=x\right] 
	&= (-1)^{k+1} \, {k+l-2 \choose k-1} \left[ 
	v_\perp^\mu(x,\tau) \, \p_\mu 
	+ \frac{1}{2} \, v_{2\perp}^{\mu\nu}(x,\tau) \, \p_\nu\p_\mu 
	\right] f(x) \qquad \forall \, k,l\geq1 \, .
\end{align}
Furthermore,
\begin{align}
	\lim_{d\tau\rightarrow 0 } \E\left[ \frac{\rd_0^{2k} f(X(\tau))}{2\,d\tau} \, \Big| \, X(\tau)=x\right] 
	&= (-1)^{k+1} {2\,k - 2 \choose k -1} \left[
	v_\perp^\mu(x,\tau) \, \p_\mu 
	+ \frac{1}{2} \, v_{2\perp}^{\mu\nu}(x,\tau) \, \p_\nu\p_\mu 
	\right] f(x) \, ,
	\nonumber\\
	\lim_{d\tau\rightarrow 0 } \E\left[ \frac{\rd_0^{2k+1} f(X(\tau))}{2\,d\tau} \, \Big| \, X(\tau)=x\right] 
	&= 0 \,  \qquad \forall \, k\geq1 \, .
\end{align}

\begin{thebibliography}{99}
	
\bibitem{Riemann:1868}
B. Riemann,
``\"Uber die Hypothesen, welche der Geometrie zu Grunde liegen.''
Abhandlungen der K\"oniglichen Gesellschaft der Wissenschaften zu G\"ottingen 13, p. 133-150 (1868).

\bibitem{Jost:2016}
J. Jost,
``Bernhard Riemann: on the hypothesis which lie at the bases of geometry,''
Classic Texts in the Sciences, Springer Switzerland (2016).
%ISBN 978-3-319-26040-2 
%ISBN 978-3-319-26042-6 (eBook)
%DOI 10.1007/978-3-319-26042-6

\bibitem{Kolmogorov}
A.~Kolmogorov,
``Grundbegriffe der Wahrscheinlichkeitsrechnung,''
Springer-Verlag (1933).

\bibitem{Schwartz:1984}
L.~Schwartz,
``Semi-Martingales and their Stochastic Calculus on Manifolds,''
Presses de l'Universit\'e de Montr\'eal (1984).

\bibitem{Meyer:1981}
P.~A.~Meyer,
``A differential geometric formalism for the It\^o calculus. Stochastic Integrals.''
Lecture Notes in Mathematics \textbf{851}, Springer (1981).

\bibitem{Feynman:1948ur}
R.~P.~Feynman,
``Space-time approach to nonrelativistic quantum mechanics,''
Rev. Mod. Phys. \textbf{20}, 367-387 (1948).
%doi:10.1103/RevModPhys.20.367

\bibitem{FKac}
M.~Kac,
``On Distribution of Certain Wiener Functionals,''
Trans. Amer. Math. Soc. \textbf{65}, 1-13 (1949).

\bibitem{Glimm:1987ylb}
J.~Glimm and A.~Jaffe,
``Quantum Physics: A Functional Integral Point of View,''
Springer-Verlag, New York (1987).
%ISBN 978-0-387-96477-5, 978-1-4612-4728-9
%doi:10.1007/978-1-4612-4728-9

\bibitem{DeWitt:1967ub}
B.~S.~DeWitt,
``Quantum Theory of Gravity. 2. The Manifestly Covariant Theory,''
Phys. Rev. \textbf{162}, 1195-1239 (1967).
%doi:10.1103/PhysRev.162.1195

\bibitem{ChapelHill}
D.~Rickles and C.M.~DeWitt,
``The Role of Gravitation in Physics: Report from the 1957 Chapel Hill Conference,''
Max-Planck-Gesellschaft zur F\"orderung der Wissenschaften, Berlin (2011).
%doi:10.34663/9783945561294-00
%ISBN:978-3-945561-29-4

\bibitem{Reisenberger:2001pk}
M.~Reisenberger and C.~Rovelli,
``Space-time states and covariant quantum theory,''
Phys. Rev. D \textbf{65}, 125016 (2002).
%doi:10.1103/PhysRevD.65.125016
%[arXiv:gr-qc/0111016 [gr-qc]].

\bibitem{Giovannetti:2022pab}
V.~Giovannetti, S.~Lloyd and L.~Maccone,
``Geometric Event-Based Quantum Mechanics,''
New J. Phys. \textbf{25}, no.2, 023027 (2023).
%doi:10.1088/1367-2630/acb793
%[arXiv:2206.08359 [quant-ph]].

\bibitem{Kuipers:2023pzm}
F.~Kuipers,
``Stochastic Mechanics: The Unification of Quantum Mechanics with Brownian Motion,''
SpringerBriefs in Physics, Spinger Cham (2023).
%ISBN 978-3-031-31447-6, 978-3-031-31448-3
%doi:10.1007/978-3-031-31448-3
%[arXiv:2301.05467 [quant-ph]].

\bibitem{Kuipers:2023ibv}
F.~Kuipers,
``Quantum mechanics from stochastic processes,''
Eur. Phys. J. Plus \textbf{138}, 542 (2023).
%doi:10.1140/epjp/s13360-023-04184-x
%[arXiv:2304.07524 [quant-ph]].

\bibitem{Cameron}
R.~H.~Cameron,
``A Family of Integrals Serving to Connect the Wiener and Feynman Integrals,''
J. Math. and Phys. \textbf{39}, p. 126-140 (1960).

\bibitem{Daletskii}
Yu.~L.~Daletskii,
``Functional integrals connected with operator evolution equations,''
Russ. Math. Surv. \textbf{17}, 5, 1-107 (1962).

\bibitem{Albeverio}
S.~A.~Albeverio, R.~J.~H{\o}egh-Krohn and S.~Mazzucchi,
``Mathematical Theory of Feynman Path Integrals,''
Lecture Notes in Mathematics \textbf{523}, Springer-Verlag (2008).

\bibitem{Wick:1954eu}
G.~C.~Wick,
``Properties of Bethe-Salpeter Wave Functions,''
Phys. Rev. \textbf{96}, 1124-1134 (1954).
%doi:10.1103/PhysRev.96.1124

\bibitem{Wiener}
N.~Wiener,
``Differential Space,''
Journal of Mathematics and Physics \textbf{2}, 131-174 (1923).

\bibitem{Morters:2010}
P.~M\"orters and Y.~Peres,
``Brownian Motion,''
Cambridge Series in Statistical and Probabilistic Mathematics, Cambridge University press, Cambridge UK (2010).

\bibitem{Nelson:1967}
E.~Nelson,
``Dynamical Theories of Brownian Motion,''
Princeton University Press (1967).

\bibitem{Isserlis:1918}
L.~Isserlis,
``On a Formula for the Product-Moment Coefficient of any Order of a Normal Frequency Distribution in any Number of Variables,''
Biometrica \textbf{12}, 134-139 (1918).
%doi:10.2307/2331932

\bibitem{Wick:1950ee}
G.~C.~Wick,
``The Evaluation of the Collision Matrix,''
Phys. Rev. \textbf{80}, 268-272 (1950).
%doi:10.1103/PhysRev.80.268


\bibitem{Chaichian:2001cz}
M.~Chaichian and A.~Demichev,
``Path integrals in physics. Vol. 1: Stochastic processes and quantum mechanics,''
CRC Press Bristol (2019).
%ISBN 9780367397142


\bibitem{Emery:1989}
M.~Emery, 
``Stochastic Calculus in Manifolds,''
Springer-Verlag (1989).

\bibitem{Huang:2022}
Q.~Huang and J.~C.~Zambrini,
``From Second-order Differential Geometry to Stochastic Geometric Mechanics,''
J. Nonlinear Sci. \textbf{33}, 67 (2023).
%arXiv:2201.03706 [math-ph] (2022).

\bibitem{Bies:2023zvs}
W.~Bies,
``Riemannian Geometry to Higher Order in the Infinitesimals,''
[arXiv:2310.04214 [math.DG]].

\bibitem{DeWitt:1957}
B.~S.~DeWitt,
``Dynamical theory in curved spaces I: a review of the classical and quantum action principles,''
Rev. Mod. Phys. \textbf{29}, 377-397 (1957). 

\bibitem{Pauli:1973}
W.~Pauli,
``Pauli Lectures on Physics 6: selected topics in field quantization,''
MIT press (1973).


\bibitem{Nelson:1985}
E.~Nelson,
``Quantum Fluctuations,''
Princeton University Press (1985).

\bibitem{Kuipers:2021ylr}
F.~Kuipers,
``Analytic continuation of stochastic mechanics,''
J. Math. Phys. \textbf{63}, 4, 042301 (2022).
%doi:10.1063/5.0073096

\bibitem{DeWitt:2003pm}
B.~S.~DeWitt,
``The global approach to quantum field theory. Vol. 1, 2,''
Int. Ser. Monogr. Phys. \textbf{114}, 1-1042 (2003).

\bibitem{DeWitt:2012mdz}
B.~S.~DeWitt,
``Supermanifolds,''
Cambridge Univ. Press (2012).
%ISBN 978-1-139-24051-2, 978-0-521-42377-9
%doi:10.1017/CBO9780511564000

\bibitem{Schuller:2005yt}
F.~P.~Schuller and M.~N.~R.~Wohlfarth,
``Geometry of manifolds with area metric: multi-metric backgrounds,''
Nucl. Phys. B \textbf{747}, 398-422 (2006).
%doi:10.1016/j.nuclphysb.2006.04.019
%[arXiv:hep-th/0508170 [hep-th]].

\bibitem{Lust:1989tj}
D.~Lust and S.~Theisen,
``Lectures on string theory,''
Lect. Notes Phys. \textbf{346}, 1-346 (1989).
%doi:10.1007/BFb0113507

\bibitem{Yasue:1981}
K.~Yasue,
``Stochastic calculus of variations,''
J. Funct. Anal. \textbf{41}, 3, 327 (1981).

\bibitem{Guerra:1982fn}
F.~Guerra and L.~M.~Morato,
``Quantization of Dynamical Systems and Stochastic Control Theory,''
Phys. Rev. D \textbf{27}, 1774 (1983).
%doi:10.1103/PhysRevD.27.1774

\bibitem{Pavon:1995April}
M.~Pavon,
``A new formulation of stochastic mechanics,''
Phys. Lett. A \textbf{209}, 143-149 (1995).

\bibitem{Pavon:2000}
M.~Pavon,
``Stochastic mechanics and the Feynman integral,''
J. Math. Phys. \textbf{41}, 6060 (2000).

\bibitem{Woodard:2015zca}
R.~P.~Woodard,
``Ostrogradsky's theorem on Hamiltonian instability,''
Scholarpedia \textbf{10}, no.8, 32243 (2015).
%doi:10.4249/scholarpedia.32243
%[arXiv:1506.02210 [hep-th]].

\bibitem{Grudka:2024llq}
A.~Grudka, J.~Oppenheim, A.~Russo and M.~Sajjad,
``Renormalisation of postquantum-classical gravity,''
[arXiv:2402.17844 [hep-th]].

\bibitem{Hitchin:2003cxu}
N.~Hitchin,
``Generalized Calabi-Yau manifolds,''
Quart. J. Math. Oxford Ser. \textbf{54}, 281-308 (2003).
%doi:10.1093/qjmath/54.3.281
%[arXiv:math/0209099 [math.DG]].

\bibitem{Gualtieri:2003dx}
M.~Gualtieri,
``Generalized complex geometry,''
Ann. Math. \textbf{174}, 75–123 (2011).
%doi: 10.4007/annals.2011.174.1.3
%[arXiv:math/0401221 [math.DG]].

\bibitem{Gualtieri:2007bq}
M.~Gualtieri,
``Branes on Poisson varieties,''
in ``The Many Facets of Geometry: A Tribute to Nigel Hitchin'' Oxford University Press (2010). 
%doi:10.1093/acprof:oso/9780199534920.003.0018
%[arXiv:0710.2719 [math.DG]].

\bibitem{Siegel:1993th}
W.~Siegel,
``Superspace duality in low-energy superstrings,''
Phys. Rev. D \textbf{48}, 2826-2837 (1993).
%doi:10.1103/PhysRevD.48.2826
%[arXiv:hep-th/9305073 [hep-th]].

\bibitem{Hull:2009mi}
C.~Hull and B.~Zwiebach,
``Double Field Theory,''
JHEP \textbf{09}, 099 (2009).
%doi:10.1088/1126-6708/2009/09/099
%[arXiv:0904.4664 [hep-th]].

\bibitem{Hull:2009zb}
C.~Hull and B.~Zwiebach,
``The Gauge algebra of double field theory and Courant brackets,''
JHEP \textbf{09}, 090 (2009).
%doi:10.1088/1126-6708/2009/09/090
%[arXiv:0908.1792 [hep-th]].

\bibitem{Hohm:2010jy}
O.~Hohm, C.~Hull and B.~Zwiebach,
``Background independent action for double field theory,''
JHEP \textbf{07}, 016 (2010).
%doi:10.1007/JHEP07(2010)016
%[arXiv:1003.5027 [hep-th]].

\bibitem{Hohm:2010pp}
O.~Hohm, C.~Hull and B.~Zwiebach,
``Generalized metric formulation of double field theory,''
JHEP \textbf{08}, 008 (2010).
%doi:10.1007/JHEP08(2010)008
%[arXiv:1006.4823 [hep-th]].

\bibitem{Hohm:2013bwa}
O.~Hohm, D.~L\"ust and B.~Zwiebach,
``The Spacetime of Double Field Theory: Review, Remarks, and Outlook,''
Fortsch. Phys. \textbf{61}, 926-966 (2013).
%doi:10.1002/prop.201300024
%[arXiv:1309.2977 [hep-th]].

\bibitem{Bies:202406}
W.~Bies,
``Post-Einsteinian Effects in the General Theory of Relativity from Higher-Order Riemannian Geometry,''
[arXiv:2406.06604 [math.DG]].

\bibitem{Bies:2024gjn}
W.~Bies,
``Unification of the Fundamental Forces in Higher-Order Riemannian Geometry,''
[arXiv:2406.06605 [math.DG]].

\bibitem{Arzano:2024apl}
M.~Arzano and F.~Kuipers,
``A Stochastic Origin of Spacetime Non-Commutativity,''
[arXiv:2409.11866 [hep-th]].

\bibitem{Tao:2016}
T. Tao,
``Higher Order Fourier Analysis,''
Graduate Studies in Mathematics Vol. 142, American Mathematical Society (2012).
%ISBN 978-1-4704-5998-7

\bibitem{Stelle:1976gc}
K.~S.~Stelle,
``Renormalization of Higher Derivative Quantum Gravity,''
Phys. Rev. D \textbf{16}, 953-969 (1977).
%doi:10.1103/PhysRevD.16.953

\bibitem{Stelle:1977ry}
K.~S.~Stelle,
``Classical Gravity with Higher Derivatives,''
Gen. Rel. Grav. \textbf{9}, 353-371 (1978).
%doi:10.1007/BF00760427

\bibitem{Schuller:2005ru}
F.~P.~Schuller and M.~N.~R.~Wohlfarth,
``Canonical differential geometry of string backgrounds,''
JHEP \textbf{02}, 059 (2006)
%doi:10.1088/1126-6708/2006/02/059
%[arXiv:hep-th/0511157 [hep-th]].

\bibitem{Punzi:2006hy}
R.~Punzi, F.~P.~Schuller and M.~N.~R.~Wohlfarth,
``Geometry for the accelerating universe,''
Phys. Rev. D \textbf{76}, 101501 (2007)
%doi:10.1103/PhysRevD.76.101501
%[arXiv:hep-th/0612133 [hep-th]].

\bibitem{Punzi:2006nx}
R.~Punzi, F.~P.~Schuller and M.~N.~R.~Wohlfarth,
``Area metric gravity and accelerating cosmology,''
JHEP \textbf{02}, 030 (2007).
%doi:10.1088/1126-6708/2007/02/030
%[arXiv:hep-th/0612141 [hep-th]].

\bibitem{Dittrich:2023ava}
B.~Dittrich and J.~Padua-Arg\"uelles,
``Twisted geometries are area-metric geometries,''
Phys. Rev. D \textbf{109}, no.2, 026002 (2024).
%doi:10.1103/PhysRevD.109.026002
%[arXiv:2302.11586 [gr-qc]].

\bibitem{Borissova:2023yxs}
J.~N.~Borissova, B.~Dittrich and K.~Krasnov,
``Area-metric gravity revisited,''
Phys. Rev. D \textbf{109}, no.12, 124035 (2024).
%doi:10.1103/PhysRevD.109.124035
%[arXiv:2312.13935 [gr-qc]].

\bibitem{Beggs:2020}
E.~J.~Beggs and S.~Majid
``Quantum Riemannian Geometry,''
Grundlehren der mathematischen Wissenschaften, Spinger Cham (2020).
%ISBN 978-3-030-30293-1, 978-3-030-30296-2, 978-3-030-30294-8
%doi:10.1007/978-3-030-30294-8


%\cite{Connes:1997cr}
\bibitem{Connes:1997cr}
A.~Connes, M.~R.~Douglas and A.~S.~Schwarz,
``Noncommutative geometry and matrix theory: Compactification on tori,''
JHEP \textbf{02}, 003 (1998).
%doi:10.1088/1126-6708/1998/02/003
%[arXiv:hep-th/9711162 [hep-th]].

\bibitem{Sheikh-Jabbari:1997qke}
M.~M.~Sheikh-Jabbari,
``More on mixed boundary conditions and D-branes bound states,''
Phys. Lett. B \textbf{425}, 48-54 (1998).
%doi:10.1016/S0370-2693(98)00199-3
%[arXiv:hep-th/9712199 [hep-th]].

%\cite{Ardalan:1998ce}
\bibitem{Ardalan:1998ce}
F.~Ardalan, H.~Arfaei and M.~M.~Sheikh-Jabbari,
``Noncommutative geometry from strings and branes,''
JHEP \textbf{02}, 016 (1999).
%doi:10.1088/1126-6708/1999/02/016
%[arXiv:hep-th/9810072 [hep-th]].

%\cite{Seiberg:1999vs}
\bibitem{Seiberg:1999vs}
N.~Seiberg and E.~Witten,
``String theory and noncommutative geometry,''
JHEP \textbf{09}, 032 (1999).
%doi:10.1088/1126-6708/1999/09/032
%[arXiv:hep-th/9908142 [hep-th]].

\bibitem{Kuipers:2021jlh}
F.~Kuipers,
``Stochastic Quantization on Lorentzian Manifolds,''
JHEP \textbf{05}, 028 (2021).
%doi:10.1007/JHEP05(2021)028
%[arXiv:2101.12552 [hep-th]].

\bibitem{Kuipers:2022wpy}
F.~Kuipers,
``Spacetime Stochasticity and Second Order Geometry,''
Springer Proc. Math. Stat. \textbf{396}, 395-400 (2022).
%doi:10.1007/978-981-19-4751-3\_35
%[arXiv:2203.16399 [gr-qc]].

\bibitem{Percacci:1984ai}
R.~Percacci,
``Spontaneous Soldering,''
Phys. Lett. B \textbf{144}, 37-40 (1984).
%doi:10.1016/0370-2693(84)90171-0

\bibitem{Percacci:1990wy}
R.~Percacci,
``The Higgs phenomenon in quantum gravity,''
Nucl. Phys. B \textbf{353}, 271-290 (1991).
%doi:10.1016/0550-3213(91)90510-5
%[arXiv:0712.3545 [hep-th]].

\bibitem{Nesti:2009kk}
F.~Nesti and R.~Percacci,
``Chirality in unified theories of gravity,''
Phys. Rev. D \textbf{81}, 025010 (2010).
%doi:10.1103/PhysRevD.81.025010
%[arXiv:0909.4537 [hep-th]].

\bibitem{Chamseddine:2010rv}
A.~H.~Chamseddine and V.~Mukhanov,
``Gravity with de Sitter and Unitary Tangent Groups,''
JHEP \textbf{03}, 033 (2010).
%doi:10.1007/JHEP03(2010)033
%[arXiv:1002.0541 [hep-th]].

\bibitem{Chamseddine:2013hwa}
A.~H.~Chamseddine and V.~Mukhanov,
``Who Ordered the Anti-de Sitter Tangent Group?,''
JHEP \textbf{11}, 095 (2013).
%doi:10.1007/JHEP11(2013)095
%[arXiv:1308.3199 [hep-th]].

\bibitem{Chamseddine:2016pkx}
A.~H.~Chamseddine and V.~Mukhanov,
``On Unification of Gravity and Gauge Interactions,''
JHEP \textbf{03}, 020 (2016).
%doi:10.1007/JHEP03(2016)020
%[arXiv:1602.02295 [hep-th]].

\bibitem{Krasnov:2017epi}
K.~Krasnov and R.~Percacci,
``Gravity and Unification: A review,''
Class. Quant. Grav. \textbf{35}, no.14, 143001 (2018).
%doi:10.1088/1361-6382/aac58d
%[arXiv:1712.03061 [hep-th]].

\bibitem{Konitopoulos:2023wst}
S.~Konitopoulos, D.~Roumelioti and G.~Zoupanos,
``Unification of Gravity and Internal Interactions,''
Fortsch. Phys. \textbf{72}, no.1, 2300226 (2024).
%doi:10.1002/prop.202300226
%[arXiv:2309.15892 [hep-th]].

\bibitem{Roumelioti:2024lvn}
D.~Roumelioti, S.~Stefas and G.~Zoupanos,
``Unification of conformal gravity and internal interactions,''
Eur. Phys. J. C \textbf{84}, no.6, 577 (2024).
%doi:10.1140/epjc/s10052-024-12949-6
%[arXiv:2403.17511 [hep-th]].

\bibitem{KaratzasSchreve}
I.~Karatzas and S.E.~Shreve,
``Brownian Motion and Stochastic Calculus''
Graduate Texts in Mathematics, vol 113. Springer, New-York (1998).
%doi:10.1007/978-1-4612-0949-2_2

\bibitem{Biane:2010}
P.~Biane,
``It\^o's stochastic calculus and Heisenberg commutation relations,'' 
Stoch. Process. Their Appl. \textbf{120}, 698–720 (2010).







\end{thebibliography}
\end{document}